\documentclass[pdflatex,sn-basic]{sn-jnl}

\usepackage{graphicx}%
\usepackage{multirow}%
\usepackage{amsmath,amssymb,amsfonts}%
\usepackage[title]{appendix}%
\usepackage{xcolor}%
\usepackage{booktabs}%

\usepackage{ulem}

\begin{document}

\title[RDA-PSO Computational Method]{RDA-PSO: A computational method to quantify the diffusive dispersal of insects}

\author*[1]{\fnm{Lidia} \sur{Mrad}}\email{lmrad@mtholyoke.edu}

\author[2]{\fnm{Joceline} \sur{Lega}}\email{lega@arizona.edu}

\affil*[1]{\orgdiv{Department of Mathematics and Statistics}, \orgname{Mount Holyoke College}, \orgaddress{\street{50 College St.}, \city{South Hadley}, \postcode{01075}, \state{MA}, \country{USA}}}

\affil[2]{\orgdiv{Department of Mathematics}, \orgname{University of Arizona}, \orgaddress{\street{617 N Santa Rita Ave.}, \city{Tucson}, \postcode{85721}, \state{AZ}, \country{USA}}}

\abstract{This article introduces a computational method, called {\em Recapture of Diffusive Agents \& Particle Swarm Optimization} (RDA-PSO), designed to estimate the dispersal parameter of diffusive insects in mark-release-recapture (MRR) field experiments. In addition to describing the method, its properties are discussed, with particular focus on robustness in estimating the observed diffusion coefficient in the presence of uncertainty. It is shown that RDA-PSO provides a simple and reliable approach to quantify insect dispersal that can handle low recapture rates and uneven capture site distributions without the need for area corrections. Tests on synthetic data, for which the actual diffusion coefficient is known, show the method outperforms three techniques based on the solution of the diffusion equation, which are also introduced in this work. Examples of application to real field data for the yellow fever mosquito are provided.}



\maketitle
\section{Introduction}
The recent epidemics of chikungunya \citep{SRL24}, Zika \citep{Fe16} and dengue \citep{Le24} in the Americas, as well as various infestations in Europe \citep{GBB23}, have dramatically increased the level of attention given to the yellow fever mosquito, \textit{Aedes aegypti}, which is the primary dengue vector. Many recent publications have highlighted the broad geographic range of areas that are suitable for this species development \citep{KSD15,CDC17}, and efforts to combat disease by releasing mutants in local populations have multiplied. Understanding the dispersal of \textit{Ae. aegypti} is therefore useful both for applications related to the controlled release of mutant males (e.g. \citep{EMC17}) and to model female abundance patterns at various scales. 

\textit{Ae. aegypti} tends to frequent inhabited areas and is typically expected not to fly further than 30 meters from a suitable household \citep{LMA14}, although in rare occasions it was observed several hundreds of meters away from its point of release \citep{RAA95}. Its infestation patterns are complex. They may be temporally stable at the neighborhood level (\citep{Ba11} cited in \citep{LMA14}), but at the same time distributions can be highly localized at the household scale, with clusters that are temporally unstable \citep{LMA14}. Because areas with elevated mosquito abundance have been associated with disease presence (e.g. dengue \citep{VCF15}) and because zones of high larval infestation are not necessarily good indicators of mosquito hotspots \citep{VCF15}, models that couple adult mosquito abundance with spatial dynamics have the potential to be useful guides for public health interventions. However, detailed information on how \textit{Ae. aegypti} mosquitoes cluster around households \citep{LMA14}, as well as on the effect of the local built on disease transmission  \citep{KA17} is lacking.

This article focuses on diffusive models of \textit{Ae. aegypti} dispersal and presents methods to measure the associated diffusion coefficient. In particular, we introduce a computational model that accurately estimates diffusive dispersal parameters associated with mark-release-recapture (MRR) data, using particle swarm optimization. In MRR experiments, a number of marked insects are released in the environment at the same time from a single location. In the following days and weeks, researchers systematically search for the released insects at regular times and record when and where they were recaptured. Typically, only a few percent of the released insects are found. We analyze the robustness of the computational approach presented here with respect to changes in experimental parameters (e.g. size and attractiveness of insect capture sites) and quantify the uncertainty associated with both experimental conditions and computational limitations. Section \ref{sec:dispersal} reviews the literature on mosquito dispersal, focusing on why a diffusive process (possibly with advection in the presence of wind or bias towards attracting sites) is a reasonable approximation. Section \ref{sec:Methods} introduces a simple calculation based on the mean distance traveled (MDT) by mosquitoes recaptured in MRR experiments, as well as three optimization-based approaches to estimating the diffusion coefficient from recapture data. The \textit{time-corrected} and \textit{area-and-time-corrected} models use theoretical expressions derived from the diffusion equation and are based on the assumption that recapturing released mosquitoes (often done by aspirating insects found in a specific location at a given time or by using traps) amounts to sampling their overall distribution. The \textit{Recapture of Diffusive Agents \& Particle Swarm Optimization} (RDA-PSO) model, which is the main focus of this article, simulates insect diffusion and recapture at the microscopic level. Section \ref{sec:Results} discusses the performance of the three optimization-based models on synthetic and real-world recapture data and numerically quantifies uncertainty associated with RDA-PSO estimates. Section \ref{sec:Discussion} summarizes the results and highlights the strengths of the proposed computational approach in estimating dispersal parameters from noisy data. In particular, the RDA-PSO method introduced here is robust to lack of knowledge on the precise location of capture sites, its input data do not have to be corrected to account for unevenly distributed recapture locations, and it works well even when the number of recaptured mosquitoes is small, which is typical of mark-release-recapture experiments. 

\section{Modeling of mosquito dispersal}
\label{sec:dispersal}
Mosquitoes typically disperse to rest, mate, feed (on nectar or on blood), and to lay eggs \citep{Se97}. Early quantitative studies of their dispersal behavior go back more than 100 years (Ross’s work \citep{Ro05,Ro11} cited by Service \citep{Se97}). Mark-release-recapture (MRR) experiments, in which tagged insects are released in the environment and recaptured at later times, have been used to measure native population abundance or compare survival rates of different morphs in a polymorphic population (see e.g. \citep{Ma85}). They also provide a way to quantify arthropod dispersal. Typically, concentric annular zones are defined about the release point and numbers of marked insects captured in each zone are reported during a period of a few days, weeks or even months after the initial release. Recapture may be done by traps or by systematic aspiration in pre-specified areas (e.g. houses). In this article, we use the words trap or capture site interchangeably. For more details, the reader is referred to \citep{Si08}. Two concepts are central to the analysis of MRR data. First, one needs to model how mosquitoes move; we will see that there is plenty of evidence to describe such dynamics in terms of regular diffusion. Second, one needs to assess how to relate recapture data to mosquito density.

\subsection{Dispersal at the microscopic level}
It is natural to describe dispersal at the level of individual insects in terms of a random walk \citep{OL01,CPB08}, consisting of successive displacements along straight lines alternating with changes in the direction (or angle with respect to an axis of reference) in which each straight forward motion occurs. The use of random walks in biology is of course very common \citep{Be93,CPB08}, from modeling how bacteria swim by alternating runs and tumbles \citep{Be93} to, at a much bigger scale, how animals look for food. In many situations, these walks are expected to be correlated \citep{KS83,ABPB21}, due to a general tendency of moving forward \citep{CPB08}: each new direction of motion is close to the preceding one, thereby limiting the likelihood of large turning angles. However, after many steps, such a random process is equivalent to an isotropic random walk, in the sense that the mean square displacement scales linearly with the number of steps \citep{KS83,PPB14,ABPB21}. Thus any local directional bias arising from correlated consecutive steps diminishes over time, and the resulting long-term dynamics may be replaced by a regular random walk, with uniformly distributed step orientations. In other words, the memory of short-term correlations is eventually lost \citep{CPB08} and, in the absence of overall bias, insect dispersal is well described by isotropic diffusion at the macroscopic level \citep{CPB08,PPB14,ABPB21}. 

In the case of MRR experiments, the above diffusive limit assumes all insects participating in the random walk behave in a similar way. In 1943, Dobzhansky \& Wright \citep{DW43} analyzed the behavior of \textit{Drosophila pseudoobscura} in four release-recapture experiments in which mutants were caught in traps arranged along two perpendicular lines that crossed at the point of release. The distribution of mutant flies at a given distance from that point was estimated from trap data and found to be leptokurdic, which raised questions about the diffusive nature of insect dispersal. In their 1943 article \citep{DW43}, Dobzhansky and Wright suggested their observations may be the result of the superposition of two normal distributions with different variances. Skellam \citep{Sk51} proposed that the presence a leptokurdic distribution could be due to differences between individual insect dispersal rates, and argued that such observations do not mean that dispersal is not random (in the sense of resulting from a Markovian random walk). It is now generally accepted that in simple situations such as MRR experiments, mosquito dispersal is essentially diffusive.

\subsection{Dispersal at the macroscopic level}
At the macroscopic level, dispersal of a large number of insects according to an isotropic random walk is described by regular diffusion. Macroscopic models of mosquito spread typically include diffusion, as well as wind dispersal \citep{TMF05}, the latter being represented by an advection term. Such a description is in line with reaction-diffusion models of biological dispersal \citep{Le76,CPB08}, which may include terms describing advection, directional bias (chemotaxis or other form of taxis), super- and sub-diffusion, nonlinear diffusion (for instance to model larger dispersal rates due to overcrowding), or space-dependent parameters. In the case of urban vectors, landscape features such as streets and buildings may be accounted for by letting parameters vary discontinuously between different sections of the computational or modeling domain \citep{YDC18,BU22,RNN22}. Reaction terms describing the growth, decay, or interactions between compartments associated with different stages of mosquito development are included to capture the long-term dynamics of these populations \citep{YDC18,SLC20,LVC23,RNN22}. Recently, vector control approaches involving the release of genetically modified mosquitoes have led to the development of reaction-diffusion models \citep{dSLP22,IFH21,CK13} with different parameters for different groups of adult mosquitoes (such as females, wild-type males, and genetically modified males \citep{dSLP22}, or females that disperse either locally or over longer distances \citep{CK13}). Models concerned with longer time scales, for instance those focusing on potential consequences of climate change \citep{BDL18}, allow parameters to be time-dependent. At larger spatial scales, diffusion-based models with space-dependent parameters and seasonality effects (leading to space- and time-dependent growth rates) have been used to describe the spread of mosquito-borne diseases \citep{FMW17}.

All of these models require quantification of diffusive dispersal in terms of an associated diffusion coefficient. The latter may be estimated by matching a solution of the relevant diffusion, reaction-diffusion, or reaction-advection-diffusion equation with appropriate initial conditions to measurements in the field \citep{SLH71,Ka83,CR96,SPDBC17}, while paying attention to possible sources of heterogeneity \citep{Ka82,SPDBC17}. Indeed, dispersal properties not only depend on the species being considered, but also on the environment in which spread occurs \citep{SPDBC17}, and possibly on the time elapsed since the dispersing insects were introduced in that environment \citep{TMBBBPS17}. It is in general accepted that initial dispersal after release of displaced insects is typically stronger than for non-displaced insects (\citep{DW43} for flies; \citep{Se97,HSL05,MCL07,MSC10} for \textit{Ae. aegypti}). Moreover, it was suggested that larger initial dispersal may be due to overcrowding \citep{Wa66}. Many MRR experiments with mosquitoes also report the existence of long-range dispersal just after release from within an area not favored by the insects, followed by shorter-range dispersal once they arrive in a suitable location \citep{VCF15,TTW01,EMC17}. For \textit{Ae. aegypti} released in urban areas or in villages, the distance between houses affects the mean distance traveled from the point of release \citep{TTW01,HSL05,MCL07}. Moreover, \textit{Ae. aegypti} rarely dispersed beyond village boundaries \citep{HSL05} in rural MRR experiments, leading to more uniform distributions of recaptured mosquitoes as time went on in small villages (for instance \citep{THC95}). In addition, insects can exhibit tendencies to aggregate in certain areas \citep{DW43}, making mosquito range clearly environment dependent (\citep{TCB13}, citing Gillies and de Meillon \citep{GM68}). An exact calculation of the diffusion coefficient associated with a random walk on a lattice with obstacles is possible \citep{EBS15}, but is only practical for small lattice sizes. 

\subsection{Estimation of dispersal-related bionomic parameters}
\label{sec:BionomicParams}
An important outcome of MRR experiments is an estimation of the mean distance traveled (MDT) by the released insects during the period of the experiment, which may be used to quantify distances over which an infected vector may spread a disease. There is a vast literature on MRR experiments (see \citep{GRP14,MB22} for mosquitoes) and it is natural to use diffusive models to understand their outcome \citep{Ka82,TT93,CR96,ABPB21}. Indeed, experiments with mosquitoes (e.g. \citep{ML09}) have shown that in the absence of natural barriers, no preferred flight direction is observed, supporting a description of dispersal in terms of a uniform random walk. In such a case, the MDT is directly related to the diffusion coefficient. Proper interpretation of MDT measurements however depends on the context of each MRR experiment. Indeed, it is known that the MDT is limited by the spatial extent of the experiment \citep{DW43}, that it is expected to change with temperature \citep{DW43} as well as location, and that its value depends on adult type (male, female, or gravid female) \citep{JGC20,TPJ21} and on seasonality \citep{JGC20}. This variability is a consequence of how the environment affects dispersal, as detailed at the end the previous section. Finally, it should be kept in mind that the MDT over the lifetime of a mosquito is expected to be larger than the MDT estimated from MRR experiments \citep{TPZ20}.

A diffusive model of dispersal from a single release point predicts a Gaussian density whose shape flattens with time. However, the question of estimating mosquito density from recapture data is not straightforward. If systematic collection in an area is performed, then it is reasonable to assume that the number of recaptured insects is proportional to their local density (or abundance per unit area) \citep{Ka82,TT93,CR96}. When recapture first requires that the insect be spotted in the environment (e.g. when manual aspiration is used), or if there is some randomness involved in the recapture process (e.g. when traps are used), the number of insects caught in a region of limited spatial extent is likely to be described by a Poisson distribution whose rate is proportional to the local insect density \citep{YME22,VMM21,TPZ20} or by a binomial distribution \citep{MRB19}. Numerical simulations have shown that very efficient trapping leads to significant depletion of insects in the vicinity of a trap \citep{PBA12,ABPB21}, that the number of captures scales like the perimeter of the trap \citep{PPB14}, and that trap shape, as well as how the step size of a simulated random walk compares to the trap size, matter \citep{ABPB21}. However, since the recapture rate of most MRR experiments is often of the order of a few percent, some of these effects are likely to be diminished. Finally, because the trap distribution is typically not uniform in space, a correction factor \citep{LMJ81,MLL91,TPZ20,HAD14} is often applied to trap counts, in order to account for lower trap density away from the release point. When recaptured numbers are low and their variance is large, such corrections amplify their variability, potentially affecting quantitative estimates of the MDT.

Wallace \citep{Wa66} remarked on the existence of an empirical linear relationship between the logarithm of the number of insects (flies in the case of Dobzhansky \& Wright’s 1943 data) recaptured in traps and either time since release or the square root of distance from the release point. A similar relationship was reported in \citep{THC95} and \citep{TTW01}, but as a linear function of distance; in \citep{THC95} (see Figure 5 of that article), the number of mosquitoes per house was also described as a decreasing linear function of the square root of distance, with slope getting closer to zero (corresponding to a more uniform distribution of released mosquitoes) as time went on. Although the above relationships do not immediately seem consistent with diffusive behavior, it is important to assess whether any discrepancy could be related to experimental conditions. Along the same lines, Thomas \citep{TCB13} showed that a negative exponential as well as a Cauchy distribution were good fits for the number of female \textit{Anopheles gambiae} captured per night as a function of distance from breeding sites, and that the resulting distribution was well simulated by a random walk in which walkers died at a constant rate. 

\subsection{Computational models of mosquito dispersion}
Diffusion-based computational models of mosquito dispersal are, broadly speaking, agent-based, PDE-based, or discrete meta-population models. Agent-based models \citep{PPB15,ZPZ16,IFB09,STG18,WLJ22,MD17,AFE10} describe dynamics at the microscopic level, by following individual insects that perform isotropic or correlated random walks. Some meta-population models \citep{OSS08,LPS13,LKR18,OEB14,HRK19,DVR22,DMV21} describe movement in a heterogeneous landscape in terms of flux or transfer rate between different patches, including regular diffusion within a patch \citep{LPS13,LKR18} or long-range dispersal due to transportation networks \citep{DVR22,DMV21}. Other applications of diffusive behavior include the use of time-dependent diffusion kernels to assess the likelihood of mosquitoes moving between unsealed water tanks in an urban environment \citep{TPP21}. 

In the following, we introduce theoretical and computational methods to estimate the diffusive properties of dispersing insects from MRR experiments. They all use the same data as input, specifically the temporal and spatial ratios defined in Equations \eqref{eq:time_pct} and \eqref{eq:sp_pct} below. By analyzing the performance of these specific methods against synthetic data, for which the actual diffusion coefficient is known, we conclude that the three approaches based on the solution of the diffusion equation, although theoretically correct, are in practice inaccurate, whereas the RDA-PSO method produces robust results that do not require additional corrections. Although diffusive processes are well understood, the novelty of our contribution resides in its ability to address small recapture rates, the non-uniform placement of recapture locations, and the uncertainty related to how mosquitoes are attracted to capture locations.

\section{Methodology} 
\label{sec:Methods}
Our goal is to estimate the diffusion coefficient associated with MRR data. As described in Section \ref{sec:dispersal}, it is reasonable to assume the motion of insects in MRR experiments is diffusive in the region of study, away from attracting zones associated with capture sites (houses or traps). We also assume that the experiments are conducted in the absence of wind strong enough to carry some of the insects away (in particular, \textit{Aedes aegypti} does not typically fly at high altitudes, although it has been observed in upper floors of high-rise buildings \citep{AHMN20}). Heterogeneity due to landscape features is restricted to those representing capture sites. We do not factor in explicit mortality rates into our assumptions, although the daily temporal ratios defined in Equation \eqref{eq:time_pct} allow for depletion of the number of tagged mosquitoes in the environment, for instance as a consequence of mortality or because they left the area of study.

In this section, we introduce four methodological approaches to calculate the diffusion coefficient: the empirical MDT-based estimate (section \ref{sec:MDT}), the time-corrected model (section \ref{sec:TCM}), the area-and-time-corrected model (section \ref{sec:ATCM}), and the RDA-PSO computational model (section \ref{sec:CompModel}). The first three stem from theoretical considerations that are based on the solution of the diffusion equation and serve as baselines against which the computational model is compared. The latter model, which we argue is better fitted to analyze MRR experiments, relies on a description of insect dynamics at the microscopic level. All methods require the same set of input data, which we present next. 

\subsection{Preliminaries and input data}
\label{sec:InputData}

We refer to the region of study as $\mathcal{R}$, and assume that mosquitoes simultaneously and independently diffuse from a central release point before being recaptured by traps located in concentric zones of width $r_{zone}$ centered on that point. More precisely, each zone in $\mathcal{R}$ is an annulus of inner radius $r_{j-1}$ and outer radius $r_j$, where $r_j = j\,r_{zone}$, $j = 1, 2, \cdots, n_z$, and $n_z$ is the total number of zones (see for instance Figure \ref{fig:traps-by-region-TrD-5} of Appendix \ref{sec:App_synth} for an illustration). The surface area of each zone increases with its index $j$. Empirical recapture ratios, defined in Section \ref{sec:AppRecapPerc}, are used as input data.

\subsubsection{Recapture probabilities associated with diffusive spread\label{sec:RecapProb}}
We first summarize well-known formulas related to two-dimensional diffusion. In particular, the probability density function for the presence of a mosquito at distance $r$ from its point of release at time $t$ is well approximated (see details in Appendix \ref{SP:diffusion}) by
\begin{equation}
P_t (r)=\frac{r}{2 D t} \exp\left(- \frac{r^2}{4 D t}\right). \label{eq:gaussian_pdf_r}
\end{equation}
The above expression may be further integrated over $r$ to obtain the probability density function $Q_t(r_a,r_b)$ for a mosquito being present between distances $r_a$ and $r_b > r_a$ from the point of release at time $t$,
\begin{equation}
    Q_t(r_a,r_b)=\int_{r_a}^{r_b} P_t(r)\, dr = \exp\left(-\frac{r_a^2}{4 D t}\right) - \exp\left(-\frac{r_b^2}{4 D t}\right).
\label{eq:Qt_12}    
\end{equation}
In addition, Equation \eqref{eq:gaussian_pdf_r} may be used to estimate the mean distance traveled (MDT) $\langle r \rangle_t$ and the mean square distance $\langle r^2 \rangle_t$ traveled by the mosquitoes from their point of release after $t$ units of time, leading to
\begin{equation}
\langle r \rangle_t = \int_0^\infty r P_t(r)\, dr = \sqrt{\pi D t}, \qquad 
\langle r^2 \rangle_t= \int_0^\infty r^2 P_t(r)\, dr = 4 D t.
\label{eq:MDT_MSDT}
\end{equation}

\subsubsection{Empirical recapture ratios\label{sec:AppRecapPerc}}
Most MRR experiments report the number of mosquitoes recaptured on pre-determined collection dates that are typically separated by a fixed period of time (we will use one day in what follows to simplify the discussion, but the relevant unit could also be a couple of days), as well as the distance from the release point at which they were recaptured. The methods presented in this article use as input two types of recapture ratios. {\em Temporal} ratios are defined as
\begin{equation}
    \tau_i= \frac{\text{number of mosquitoes recaptured on day $i$}}{\text{total number of mosquitoes recaptured}}.
\label{eq:time_pct}
\end{equation}
and {\em spatial} ratios are defined by
\begin{equation}
    \sigma_j= \frac{\text{number of mosquitoes recaptured in zone $j$}}{\text{total number of mosquitoes recaptured}},
\label{eq:sp_pct}
\end{equation}
where, as previously mentioned, recapture zones are annular regions of width $r_{zone}$ centered on the release point. One advantage of using ratios is that if only a (constant) fraction $K$ of the mosquitoes present in the environment is captured, both $\tau_i$ and $\sigma_j$ are independent of $K$.

With the above context in place, we now introduce the four methodological approaches mentioned above, which can be applied to estimate the diffusion coefficient. Estimates obtained from the first three methods will be used as baselines for comparison with RDA-PSO estimates.

\subsection{MDT-based model}
\label{sec:MDT}

We start with a practical estimate of the diffusion coefficient based on the mean distance traveled (MDT). As noted in section \ref{sec:BionomicParams}, the empirical MDT is often reported in the MRR literature, or can easily be estimated from those reports (see for instance \citep{MB22}). Because results based on a few dozens of mosquitoes are unlikely to give reliable or consistent estimates from day to day, it is essential to aggregate mosquito counts obtained over different days in order to improve the estimate, since data from a larger number of walkers are then used to arrive at a single number. To this end, we define the MDT of an entire experiment, MDT$_{tot}$, as a weighted average of the daily MDT’s, denoted by MDT$_{i}$ for day $i$, and apply the first equation of \eqref{eq:MDT_MSDT}:  
\begin{align}
    MDT_{tot} & = \sum_{i=1}^n \tau_i\, MDT_i  = \sum_{i=1}^n \tau_i\, \sqrt{\pi D t_i}  = \sqrt{\pi D} \sum_{i=1}^n \tau_i \sqrt{t_i},
    \label{eq:MDT_tot}
\end{align}
where $n$ is the number of days mosquitoes were collected, $\tau_i$ is the temporal ratio corresponding to day $i$ defined in \eqref{eq:time_pct} and $t_i$ is the number of days elapsed since release. We emphasize that in Equation \eqref{eq:MDT_tot}, the quantity $MDT_{tot}$ defines an overall MDT, calculated for all recaptured mosquitoes, regardless of the day of recapture. This quantity is expected to be close to the empirical MDT, $MDT_{emp}$, reported in the literature (e.g. as defined in \citep{LMJ81}). Indeed,
\begin{align*}
MDT_{emp} & = \frac{1}{N_{tot}} \sum_{i=1}^n \sum_{k=1}^{N_i} r_k = \frac{1}{N_{tot}} \sum_{i=1}^n N_i \sum_{k=1}^{N_i} \frac{r_k}{N_i} \simeq \frac{1}{N_{tot}} \sum_{i=1}^n N_i\, MDT_i \\
& \simeq \sum_{i=1}^n \tau_i\, MDT_i = MDT_{tot}.
\end{align*}
In the above, $N_{tot} = \sum_{i=1}^n N_i$ is the total number of recaptured mosquitoes, $N_i$ is the number of mosquitoes captured in day $i$, and $r_k$ is the distance from the release point where the $k$-th mosquito recaptured on day $i$ was caught. The difference between $MDT_{emp}$ and $MDT_{tot}$ is that the exact expression of $MDT_i$ is used in the latter, instead of its empirical approximation $\sum_{k=1}^{N_i} \frac{r_k}{N_i}$ used in the former. However, assuming that the two values are close, knowledge of the empirical MDT, along with equation (\ref{eq:MDT_tot}), leads to the following estimate of the diffusion coefficient,
\begin{equation}
\label{eq:D_MDT}
D_{MDT} = \frac{1}{\pi} \left(\frac{MDT_{emp}}{\sum_{i=1}^n \tau_i \sqrt t_i}\right)^2,
\end{equation}
where $\tau_i$ is the temporal ratio for day $i$ and $t_i$ is the number of days since release.

\subsection{Time-corrected model\label{sec:TCM}}
One of the main difficulties of translating MRR data into estimates of dispersal parameters is that only a very small percentage of released mosquitoes is eventually recaptured. To improve empirical estimates of the quantities defined in Section \ref{sec:RecapProb}, it is therefore useful to combine information associated with different collection dates. To this end, we estimate the probability density function for a mosquito being present at a distance $r$ from the point of release at any time during the entire recapture experiment,
\[
P(r) = \int_0^\infty P_t(r) dt,
\]
as a weighted average of the probabilities $P_t(r)$ defined in Equation \eqref{eq:gaussian_pdf_r}, where the weights reflect the fraction of mosquitoes recaptured each day $i$. Thus, we  set $P(r) \simeq P_{tot}(r)$, with
\begin{equation}
    P_{tot}(r) = \sum_{i=1}^n \tau_i \, P_{t_i}(r), \quad \tau_i= \text{temporal ratios defined in Equation } \eqref{eq:time_pct},
\label{eq:P_tot}
\end{equation}
where $n$ is the total number of days mosquitoes were collected. To understand why weighting each term by $\tau_i$ is appropriate, first consider the case where there is no mosquito depletion. In such a situation, if we assume that recapture is proportional to presence, the number of mosquitoes captured on day $i$ does not depend on $t_i$, since $\int_0^\infty P_t(r) dr = 1$ for all $t$. This implies that $\tau_i = 1/n$. Consequently, the total number of mosquitoes collected over $n$ days is
\[
K N \int_0^\infty \left(\sum_{i=1}^n \tau_i \, P_{t_i}(r)\right) dr = K N \sum_{i=1}^n \left(\frac{1}{n} \int_0^\infty P_t(r) dr \right) = K N,
\]
where $N$ is the number of mosquitoes initially released and $K$ measures the total recapture ratio. However, in the presence of depletion, due to mosquitoes being caught, flying out of the study zone, or dying, the number captured on day $i$ is $K N_i$, where $N_i$ is close to but slightly less than $N$, which leads to $\tau_i = N_i / \left(\sum_{i=1}^n N_i\right)$. The definition of $P_{tot}$ is such that
\[
\int_0^\infty P_{tot}(r) dr = \int_0^\infty \left( \sum_{i=1}^n \frac{N_i}{\sum_{i=1}^n N_i} \, P_{t_i}(r) \right) dr  = 1,
\]
while taking into account that, as the MRR experiment goes on, the number of mosquitoes remaining to be captured slowly decreases.

A similar weighted average may be calculated for the probability of a mosquito being present in an annular zone between distances $r_{j-1}$ and $r_j$ from the point of release, resulting in the following expression.
\begin{align}
    Q_t(r_{j-1},r_j) & \simeq Q_{tot}(r_{j-1},r_j)\nonumber\\
    & = \sum_{i=1}^n \tau_i\, Q_{t_i}(r_{j-1},r_j), \quad \tau_i= \text{temporal ratios defined in } \eqref{eq:time_pct}.
\label{eq:Q_tot_12}
\end{align}

Estimates of $Q_{tot}$ (and $P_{tot}$) may be compared directly to data observed in MRR experiments, and an estimate of the diffusion coefficient $D$ may be obtained by optimizing the fit between model and experiment. We therefore define the {\em time-corrected model},
\begin{equation}
    Q_{tot}(r_{j-1},r_j) = \sum_{i=1}^n \tau_i\, \left[\exp\left(-\frac{r_{j-1}^2}{4 D t_i}\right) - \exp\left(-\frac{r_j^2}{4 D t_i}\right)\right],
\label{eq:Q_tot_model}   
\end{equation}
for $j$ between 1 and $n_z$, obtained by combining Equations (\ref{eq:Qt_12}) and (\ref{eq:Q_tot_12}). Because the spatial ratios $\sigma_j$, defined in equation \eqref{eq:sp_pct}, provide good approximations of $Q_{tot}(r_{j-1},r_j)$, we can estimate the diffusion coefficient $D$ by simultaneously fitting the left- and right-hand sides of the equation 
\begin{equation}
    \sigma_j = \sum_{i=1}^n \tau_i\, \left[\exp\left(-\frac{r_{j-1}^2}{4 D t_i}\right) - \exp\left(-\frac{r_j^2}{4 D t_i}\right)\right],
\label{eq:Q_tot_model_fit}    
\end{equation}
for $j$ between 1 and $n_z$. Specifically, we use Matlab's {\tt fminsearch} to minimize the sum of the squares of the differences between the left- and right-hand sides of Equation \eqref{eq:Q_tot_model_fit}. We denote the resulting estimate of the diffusion coefficient by $D_{TC}$.

\subsection{Area-and-time-corrected model} \label{sec:ATCM}
The time-corrected model discussed above implicitly assumes that capture sites are uniformly distributed across the collection area and, as previously mentioned, that the chance of capturing a mosquito at a given location is directly proportional to the likelihood of its presence. However, capture sites are not typically uniformly distributed in MRR experiments, since the number of traps in each zone is often not proportional to its surface area. Under these circumstances, it is traditional to use the correction factor introduced by Lillie et al. \citep{LMJ81}. This factor, which multiplies the original capture numbers for each trap, is zone-dependent and defined for zone $j$ by
\begin{equation}
CF_j = \frac{A_j}{A_{tot}} \, nT_{tot},
\label{eq:CF}
\end{equation}
where $A_j$ is the surface area of zone $j$, $A_{tot}$ is the total area of the circular region $\mathcal R$ where the experiment is performed, and $nT_{tot}$ is the total number of traps placed in $\mathcal R$. In other words, the corrected number of mosquitoes captured in zone $j$ is given by 
\[
N_{j}^c = CF_j \frac{N_{j}}{nT_j} = N_{j} \frac{nT_{tot}}{nT_j} \, \frac{A_j}{A_{tot}},
\]
where $N_{j}$ is the number of mosquitoes actually captured in zone $j$, and $nT_j\neq 0$ is the number of traps placed in zone $j$. The above equation can be recast as $N_j^c = N_j \cdot \rho_{tot}/\rho_j$, with the symbol $\rho$ denoting the density of traps, which shows how the correction compensates for uneven trap density. Such a correction changes the number of mosquitoes recaptured in each zone every day of the experiment and thus affects the spatial and temporal ratios defined in Equations \eqref{eq:sp_pct} and \eqref{eq:time_pct}. Fitting these corrected ratios into \eqref{eq:Q_tot_model_fit} leads to a corrected estimate of the diffusion coefficient, which we refer to as $D_{ATC}$. 

\subsection{Recapture of Diffusive Agents - Particle Swarm Optimization (RDA-PSO) model\label{sec:CompModel}}
The computational approach developed in this article provides a robust and accurate way of estimating the diffusion coefficient $D$ from temporal and spatial ratios associated with MRR data. The model is agent-based, treating mosquitoes as individual agents that perform a two-dimensional random walk in a region that mimics the conditions of the MRR experiment. Walkers do not naturally die (except when captured), but are allowed to leave the computational box, after which they do not return to the area of study. The number of traps (we use the words trap or capture site interchangeably) in each zone around the release point is assumed to be known. The estimated value of $D$ is obtained via an optimization method that simultaneously fits the spatial (see Equation \eqref{eq:sp_pct}) and temporal (see Equation \eqref{eq:time_pct}) recapture ratios. Specifically, the method aims to identify the parameters of the random walk that minimize the error $E$ between observed and simulated ratios, where $E$ is given in Equation \eqref{eq:Error2} below. There are four such parameters:
\begin{enumerate}
\item The length of each step taken by a random walker, ${\tt k} \cdot {\tt h}$ , where {\tt k} is a proportionality coefficient and {\tt h}, the mesh size of the simulation grid, serves as a reference length.
\item The size ${\tt q} \cdot {\tt h}$ of the collection region around the center of each capture site.
\item The size ${\tt p} \cdot {\tt h}$ of the attracting region around each collection region.
\item The efficiency ${\tt s}_{\tt e}$ of each capture site.
\end{enumerate}
The first parameter, {\tt k}, is directly related to the diffusion coefficient, as shown below, while the last three parameters encapsulate properties of capture sites. For simplicity, the latter are simulated as two concentric regions: the collection region is an inner disk of radius  ${\tt q} \cdot {\tt h}$ in which walkers are captured, and the attracting region is an outer annulus of width ${\tt p} \cdot {\tt h}$, in which walkers are attracted toward the center of the inner disk. Whether or not a walker feels a nearby capture site is decided, for each step taken in the disk of radius ${\tt (p+q)} \cdot {\tt h}$ around the center of the trap, by a Bernoulli draw with probability of success ${\tt s}_{\tt e}$, equal to the efficiency of the associated trap (see Appendix \ref{sec:RRDA} for details). 

For a uniform random walk, the diffusion coefficient is related to the step size  ${\tt k} \cdot {\tt h}$  by
\begin{equation}
\label{eq:DTrue_k}
    D = \frac{S}{4}({\tt k} \cdot {\tt h})^2,
\end{equation}
where $S$ is the number of steps each agent takes per unit of time (e.g. one or two days). This result may be inferred from the second equation in (\ref{eq:MDT_MSDT}), which relates $D$ to the square of the average distance traveled after $t$ units of time. Indeed, for $t = 1$, Equation (\ref{eq:MDT_MSDT}) gives $\langle r^2 \rangle_1 = 4 D$. Independently, since the steps of the random walk are not correlated, we have
\[
\langle r^2 \rangle_1 = \text{number of steps traveled per unit of time}\cdot (\text{length of each step})^2 = S \cdot ({\tt k} \cdot {\tt h})^2.
\]
Setting $4 D = S \cdot ({\tt kh})^2$ leads to the above expression for $D$. Finding {\tt k} is therefore equivalent to estimating $D$. We checked that changing $S$ in the RDA-PSO numerical method leads to changes in the optimal {\tt k} that leave the estimated $D$ unchanged, thereby allowing us to set $S$ to the reasonable but arbitrary value of 12 steps per day (see Section \ref{sec:RDA-PSO_implementation}).

Given this context, the numerical estimation proceeds as follows. For specified values of {\tt k}, {\tt p}, {\tt q}, and ${\tt s}_{\tt e}$, solving the {\em forward problem} produces an associated set of spatial and temporal ratios. This is done by simulating the random walk of $N$ agents that move according to the given parameter {\tt k} over a region with traps defined by the parameters {\tt p}, {\tt q}, and ${\tt s}_{\tt e}$, and averaging the results over 5 independent simulations. Details of the forward problem are provided in Appendix \ref{sec:App_ModelPhases}. Finding the optimal parameter values amounts to solving an {\em inverse problem}, which minimizes the error 
\begin{align}    
    E & = \frac{1}{2}(E_{spatial} + E_{temporal}) &   \nonumber  \\
    & = \frac{1}{2}\left(\sqrt{(\sigma_1-\sigma_1^o)^2+\dots+(\sigma_m-\sigma_m^o)^2} + \sqrt{(\tau_1-\tau_1^o)^2+\dots+ (\tau_n-\tau_n^o)^2}\right),
    \label{eq:Error2}
\end{align}
where $\{\sigma_1, \dots, \sigma_m\}$ are the averaged simulated spatial ratios, $\{\tau_1, \dots, \tau_n\}$ are the averaged simulated temporal ratios, $\{\sigma_1^o, \dots, \sigma_m^o\}$ are the observed spatial ratios, and $\{\tau_1^o, \dots, \tau_n^o\}$ are the observed temporal ratios. The inverse problem is solved in two steps: a grid search in the parameter space to find a suitable initial parameter range for {\tt k} and {\tt q}, followed by a particle swarm optimization (PSO) to find the parameter values that minimize $E$. The code allows the user to keep some parameters fixed (e.g. setting {\tt p} = 0 and/or $s_e = 100$~\%) and perform the optimization on the remaining parameters.

\subsubsection{Grid search}
The purpose of the grid search is to obtain an overall description of the error landscape over a suitable region of the two-dimensional parameter space for {\tt k} and {\tt q}, with {\tt p} = 0 and ${\tt s}_{\tt e} = 100\%$. The forward problem is run at specific, uniformly distributed values of {\tt k} and {\tt q}, at which the error in Equation \eqref{eq:Error2} is calculated. An example of error landscape associated with synthetic data is provided in Appendix \ref{sec:App_synth}. This error landscape clearly identifies a global minimum (see Figure \ref{fig:virtual-e2}), which corresponds to the known {\tt k} and {\tt q} parameters. Similarly, in the case of field data, a global minimum was always observed in the examples we considered, as shown in Appendices \ref{sec:App_Hainan} and \ref{sec:App_Cairns}.

However, to mimic real-world uncertainties and include the effect of the other two parameters when analyzing field experiments, we do not rely on this straightforward search method and instead develop a better optimization technique, as shown below. Specifically, the grid search serves to identify a smaller parameter range, in which the global minimum error is expected to occur, and provides the initial {\tt k} and {\tt q} range for the PSO. The latter is then performed with {\tt p} ranging from 0 to the maximum allowable size so that traps do not overlap, and ${\tt s}_{\tt e}$ ranging from 0.1\% to 100\%. 

\subsubsection{Particle swarm optimization} \label{sec:PSO}
This method finds a minimizer of the error $E$, given in equation (\ref{eq:Error2}), by iteratively moving a collection (or swarm) of $N_p = 36$ particles over the ({\tt k}, {\tt q}, {\tt p}, ${\tt s}_{\tt e}$) parameter space (see references \citep{KE95, SE98} and Appendix \ref{sec:App_PSO} for a general description of the PSO method). We use $N_g = 12$ generations, where a generation corresponds to updating the positions of all the $N_p$ particles once. For each generation, the forward problem is run at the location of each particle (in the 4-dimensional parameter space) in order to evaluate the error $E$. Each particle is then relocated, based on its own best estimation of the minimizer of the error, as well as on the best global estimation of the entire swarm. The reader is referred to Appendix \ref{sec:App_PSO} for details on how the particle positions are updated. The location of the initial group of particles (corresponding to the zeroth generation) is randomly chosen over the region where the error was found to be the smallest by the grid-search. The combination of local and global estimates typically guarantees that the particles do not gather near a local minimizer. At the end of this process, the swarm estimate of the optimal 4-tuple ({\tt k}, {\tt q}, {\tt p}, ${\tt s}_{\tt e}$) is returned. Using Equation \eqref{eq:DTrue_k}, the optimal value of {\tt k} is combined with the known values of $S$ and {\tt h} to estimate the diffusion coefficient $D$. The resulting value is referred to as $D_{RDA}$.

\subsection{Implementation of the computational model}
\label{sec:RDA-PSO_implementation}
The RDA-PSO method is designed to be implemented on realistic settings representing actual MRR experiments. To make this possible, the forward problem is run at scale, with a value of {\tt h} set at 0.1 meter. By default, mosquitoes are assumed to be released from a single point and recaptured in concentric zones around that point. The size of, and the number of capture sites contained in, each zone need to be provided. Capture sites may be traps (which typically only catch a small fraction of mosquitoes flying by) or houses, where mosquitoes are systematically collected with an aspirator. Appendix \ref{sec:App_ModelPhases} provides more details on the image processing techniques used to digitize the study area, on the implementation of the random walk, including the locally biased motion in the attracting region of each site, and on the data collection. 

In the forward problem, the motion of each agent is based on the following considerations. To take into account resting times, mosquitoes are assumed to be moving for 12 hours per day. In addition, each step of a simulated random walk represents movement over one hour. This choice is consistent with the result that many steps of a correlated random walk are well described by isotropic diffusion \citep{CPB08,PPB14,ABPB21}. Consequently, agents take $S = 12$ steps per day.

The robustness of the model is analyzed via virtual MRR experiments, where the number of walkers is set to $N = 10,000$ regardless of trap efficiency. We have checked that this leads to adequate capture rates in the outer zones, away from the release point. Indeed, there is a legitimate concern that if a large fraction of the agents are captured in zones closer to the release point, too few would make it to the outer zones, leading to capture ratios that are not representative of the actual diffusion process. Our simulations showed the ratios are stable (meaning they are the same up to a small amount of noise) for values of $N$ between 200 and 200,000. When simulating field experiments, the number of walkers is taken to be the same as the number of released mosquitoes.

Table \ref{tab:compSetup} below summarizes the computational setup and the data needed for the PSO implementation. 

\begin{table}[hbtp]
\centering
    \caption{RDA-PSO parameters. The default parameters may be adjusted by the user. The experimental constants/data depend on the field experiment that is being analyzed. The PSO-configuration parameters are set as a compromise between computational cost and result accuracy. See Appendix \ref{sec:App_PSO} for further information on the PSO configuration.}
    \label{tab:compSetup}
    \begin{tabular}{lll}
        \toprule%
        & parameter & label/value \\
        \midrule
        Default parameters & mesh size of the physical landscape & {\tt h} = 0.1 (in m) \\
        & number of steps agents take per unit time & S = 12 \\
        \midrule
        Experimental constants/data & number of mosquitoes released & $N$ \\
        & number of collection days & $n$ \\
        & width of each concentric zone & $r_{zone}$ (from maps) \\
        & release and capture site locations & (from maps)\\
        & recapture ratios & $\tau_i$ and $\sigma_j$\\
        \midrule 
        PSO-configuration parameters & number of swarming particles& $N_p= 36$\\  
        & number of generations & $N_g= 12$\\       
        \botrule%
    \end{tabular}
\end{table}

\section{Results}
\label{sec:Results}
We now establish properties of the capture ratios and compare estimates of $D$ obtained from the four models introduced in the previous section, using both synthetic and real-life MRR experiments. We also discuss limitations of the time-corrected and area-and-time-corrected models, and quantify uncertainty on estimates of $D_{RDA}$.

\subsection{Robustness of recapture ratios}
The models discussed in this article use recapture ratios, defined in Section \ref{sec:AppRecapPerc}, as their input. We analyzed the robustness of these data via multiple runs of the forward model. As expected, recapture ratios vary with trap density. They are however independent from the number $N$ of mosquito released, as long as the latter is larger than a few hundred. They are robust to changes in trap efficiency if the trap density is low, which is the case for most real-world MRR experiments. For high trap density however, the recapture ratios depend on the value of {\tt s}$_{\tt e}$. For example, if traps are uniformly distributed over the region $\mathcal{R}$ and there is one trap in the first zone (of radius $15$ m), the resulting recapture ratios are similar regardless of trap efficiency. However, if the density of traps is multiplied by 4, recapture ratios vary significantly with the chosen trap efficiency {\tt s}$_{\tt e}$. Details are provided in Appendix \ref{SP:TrapDensity}, where we remark that the latter situation would correspond to a badly designed MRR experiment, since many of the traps located in the outer zones would not catch any mosquitoes at all, and the initial pool of mosquitoes would be essentially exhausted after a few days.  Finally,  up to a small uncertainty due to the randomness of each experiment, recapture ratios appear to be independent of the exact location of the traps, as long as the number of traps in each collection zone is high enough and the trap efficiency remains low. 

\subsection{Properties of TC and ATC estimates}
Appendix \ref{SP:ATCM_prop} reports on a systematic analysis of the performance of the time-corrected and area-and-time-corrected models, via a range of synthetic MRR experiments (for which we know the exact value of {\tt k}). We consider situations where traps are evenly (leading to $D_{TC} = D_{ATC}$) or heterogeneously (where $D_{ATC}$ is relevant) distributed. When recapture ratios are robust against changes in trap location (while keeping the number of traps in each zone unchanged), the corresponding estimates of $D_{TC}$ and $D_{ATC}$ are also robust.
In addition, these estimates become more accurate when temporal percentages are evaluated hourly (i.e. after each step of the random walk) instead of daily (which in the present case corresponds to every 12 steps). Some of these results are summarized in Table \ref{tab:SyntheticD}, for uniform trap distributions. Results for non-uniformly distributed traps are presented in Appendix \ref{sec:DATCM_sen}, where it is shown that, on average, estimates of $D_{ATC}$, based on corrected temporal and spatial ratios and hourly data collection, tend to underestimate the known value $D$. 

\begin{table}[hbtp]
\centering
    \caption{Diffusion coefficient with different temporal resolutions of MRR data. Traps are uniformly distributed over the region $\mathcal{R}$, so $D_{TC} = D_{ATC}$. Low trap density corresponds to 1 trap in the first zone, while high trap density corresponds to 4 traps. Trap efficiency is set at 3\%. More setup information is provided in Appendix \ref{SP:ATCM_prop}, and values listed are part of the data used to obtain figure \ref{fig:DTC_randomization}.}
    \label{tab:SyntheticD}
    \begin{tabular}{llccc}
        \toprule%
          & & exact $D$  & $D_{ATC} (= D_{TC})$ &  [min, max] \\
          & & (m$^2/$day) & (m$^2/$day) & (m$^2/$day)\\
        \midrule
        low trap density & daily capture & \multirow{4}{*}{36.75} & 33 & [22.21, 41.3] \\
          & hourly capture &  & 36.29 & [24.65, 42.82]\\
        high trap density & daily capture &  & 30.79 &  [26.93, 33.75]\\
        & hourly capture &  & 35.59 & [32.31,  38.42]\\
        \botrule%
    \end{tabular}
\end{table}

Regardless of trap distribution, estimates of $D$ are affected by the temporal resolution (daily vs. hourly) of MRR data. This is problematic since frequent sampling of traps or systematic insect aspiration inside collection houses is time-consuming. Moreover, as previously indicated, MRR experiments described in the literature report numbers of captured mosquitoes daily or less frequently. These shortcomings clearly limit the reliability of TC and ATC estimates. As documented below, the RDA-PSO method introduced in the present work is not hindered by such considerations.

\subsection{Properties of RDA-PSO estimates}
\label{sec:Real_MRR}
The RDA-PSO inverse method estimates the four parameter {\tt k}, {\tt q}, {\tt p}, and ${\tt s}_{\tt e}$ from uncorrected temporal and spatial ratios (i.e. as defined in Equations \eqref{eq:time_pct} and \eqref{eq:sp_pct} without adjustments). Estimates of the diffusion coefficient $D$ depend directly on estimates of {\tt k}, as shown in equation (\ref{eq:DTrue_k}).  Uncertainty in the produced estimates results from arbitrariness of the initial conditions in parameter space for the zero-generation particles, from the stochastic nature of the forward model applied at each step of the method, as well as from the use of a finite number of generations. We confirmed that when applied to temporal and spatial ratios collected from virtual experiments, in which walkers spread with a known diffusion coefficient $D$, the RDA-PSO method recovers the actual value of {\tt k} and thus of $D$ with high accuracy. See Appendices \ref{sec:App_synth} and \ref{SP:ATCM_prop} for two examples. The remaining three parameters, {\tt q}, {\tt p}, and ${\tt s}_{\tt e}$, are all features of the capture sites. Our numerical explorations indicate that these parameters are not individually identifiable and that their RDA-PSO estimates adjust themselves relative to one another in predictable ways.  Importantly, we note that each estimation of the triplet ({\tt q}, {\tt p}, and ${\tt s}_{\tt e}$) produced by the RDA-PSO method is associated with a reliable estimate of the diffusion coefficient, which is therefore identifiable.

\subsubsection{Field experiments}
To illustrate the applicability of the RDA-PSO method to real-life MRR data, we now consider two MRR experiments carried out in a small village on Hainan Island in China \citep{TTW01} (we only consider the female cohorts from this study), and a third run in a suburban area of the city of Cairns in Australia \citep{RWW05}. 

In the Hainan MRR experiment \citep{TTW01}, female mosquitoes were released at two different sites: one near the center and one at the periphery of the village. Details, including the village layout, are provided in Appendix \ref{sec:App_Hainan}. The area covered by the Cairns MRR experiment is much larger (see Appendix \ref{sec:App_Cairns}), with $r_{zone} =$~ 50 meters instead of 15 meters in Hainan. Reported recapture ratios for all three release sites are shown in Tables \ref{tab:capture-table-temporal} (temporal ratios) and \ref{tab:capture-table-spatial} (spatial ratios).

\begin{table}[hbtp]
    \caption{Numbers of recaptured female mosquitoes and corresponding temporal ratios for the Hainan and Cairns studies. Left: Hainan  \citep{TTW01} when the release site was near the center of the village. Middle: Hainan \citep{TTW01} when the release site was near edge of the village. Right: Cairns \citep{RWW05}.}
    \label{tab:capture-table-temporal}
    \begin{tabular*}{\textwidth}{@{\extracolsep\fill}c@{\extracolsep\fill}c@{\extracolsep\fill}c@{\extracolsep\fill}c@{\extracolsep\fill}c@{\extracolsep\fill}c}
    \toprule%
    \multicolumn{2}{@{}c@{}}{Hainan - center} & \multicolumn{2}{@{}c@{}}{Hainan - edge} & \multicolumn{2}{@{}c@{}}{Cairns} \\\cmidrule{1-2}\cmidrule{3-4}\cmidrule{5-6}%
    Day &   Number & Day &   Number & Day &   Number \\
    &  Recaptured ($\tau_i$) &  & Recaptured ($\tau_i$) & & Recaptured ($\tau_i$) \\
    \cmidrule{1-2}\cmidrule{3-4}\cmidrule{5-6}%
        1 & 48 (45.72 \%) & 1 & 29 (38.16\%) & & \\
        2 & 14 (13.33 \%) & 2 & 12 (15.79 \%) & 5 & 20 (38.46\%)\\
        3 & 10 (9.52 \%) & 3 & 7 (9.21\%)& 8 & 18 (34.62\%)\\
        4 & 16 (15.24 \%) & 4 & 9 (11.84\%) & 11 & 9 (17.31\%)\\
        5 & 9 (8.57 \%) & 5 & 9 (11.84\%) & 15 & 5 (9.61\%)\\
        6 & 8 (7.62 \%) & 6 & 10 (13.16\%) & & \\
    \botrule
    \end{tabular*}
\end{table}

\begin{table}[hbtp]
    \caption{Numbers of recaptured female mosquitoes and corresponding spatial ratios for the Hainan and Cairns studies. Left: Hainan \citep{TTW01} when the release site was near the center of the village. Middle: Hainan \citep{TTW01} when the release site was near edge of the village, with the sixth zone boundary adjusted to exclude the furthest house, since it is assumed it did not capture any mosquitoes. Right: Cairns \citep{RWW05}.}
    \label{tab:capture-table-spatial}
    \begin{tabular*}{\textwidth}{@{\extracolsep\fill}c@{\extracolsep\fill}c@{\extracolsep\fill}c@{\extracolsep\fill}c@{\extracolsep\fill}c@{\extracolsep\fill}c}
    \toprule%
    \multicolumn{2}{@{}c@{}}{Hainan - center} & \multicolumn{2}{@{}c@{}}{Hainan - edge} & \multicolumn{2}{@{}c@{}}{Cairns} \\\cmidrule{1-2}\cmidrule{3-4}\cmidrule{5-6}%
    Zone &   Number & Zone &   Number & Zone &   Number \\
    (in m) &  Recaptured ($\sigma_j$) & (in m) &  Recaptured ($\sigma_j$) & (in m) &  Recaptured ($\sigma_j$)\\
    \cmidrule{1-2}\cmidrule{3-4}\cmidrule{5-6}%
        $(0, 15)$ & 85 (80.95\%) & $(0, 15)$ & - (0\%)\\
        $(15, 30) $ & 13 (12.38 \%) & $(15, 27) $ & 49 (64.47\%) & $(10, 60) $ & 24 (46.15\%)\\
        $(30, 45)$ & 3 (2.86\%) & $(27, 37)$ & 17 (22.37\%) & $(60, 110)$ & 16 (30.77\%)\\
        $(45, 60)$ & 4 (3.81 \%) & $(37, 60)$ & 6 (7.9\%) & $(110, 160)$ & 7 (13.46\%)\\
        $(60, 75)$ & - (0 \%) & $(60, 75) $ & 2 (2.63\%)  & $(160, 210) $ & 5 (9.62\%)\\
        $(75, 90)$ & 0 (0 \%) & $(75, 92)$ & 2 (2.63\%) & & \\
    \botrule
    \end{tabular*}
\end{table}

\subsubsection{Parameter estimation}
\label{sec:main_param_estimation}
\begin{figure}[hbtp]
    \includegraphics[width=\textwidth]{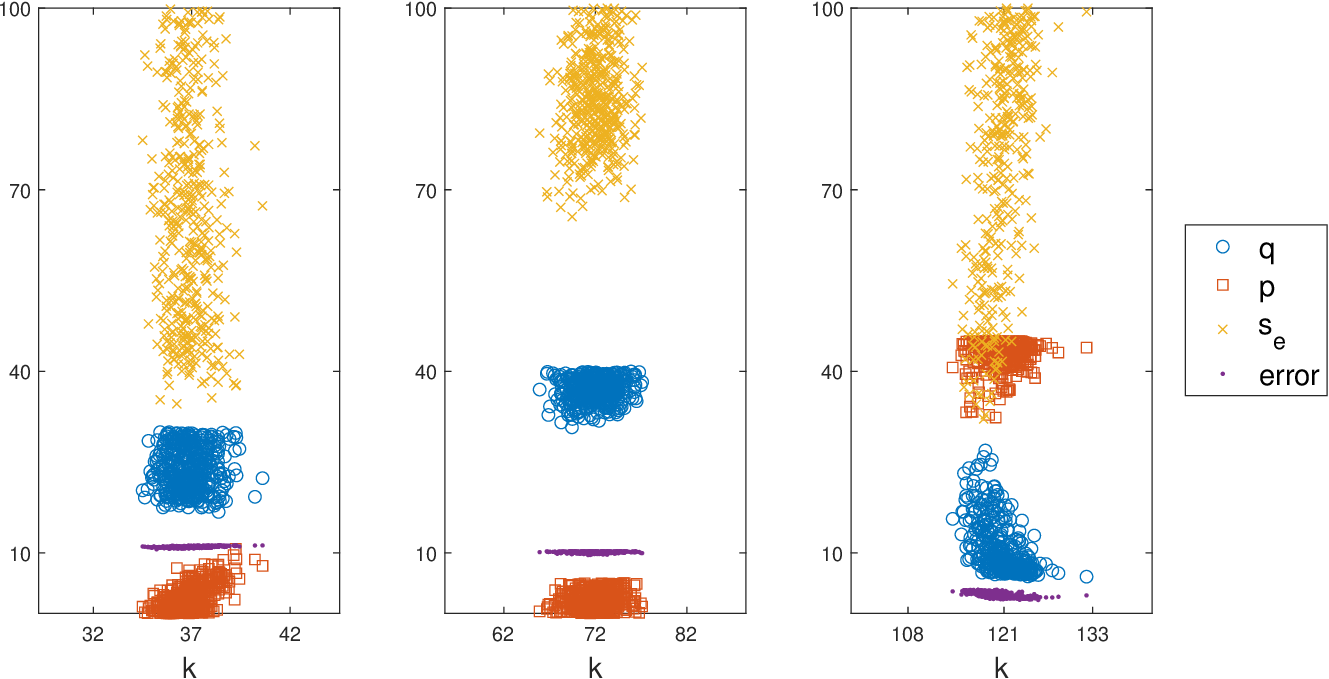}
    \caption{Results of the RDA-PSO method with tables \ref{tab:capture-table-temporal} and \ref{tab:capture-table-spatial} as input. Each returned optimal parameter 4-tuple, $({\tt k}, {\tt q}, {\tt p}, {\tt s}_{\tt e})$, is represented as a set of 3 points, $({\tt k}, {\tt q})$ (blue circles), $({\tt k}, {\tt p})$ (red squares), and $({\tt k}, {\tt s}_{\tt e})$ (yellow multiplication signs). The corresponding error is also plotted as a function of {\tt k} (purple dots). Out of 500 parameter tuples obtained, those associated with an error one half standard deviation (Cairns) or one standard deviation (Hainan) above the mean are considered outliers and were removed. Left panel: Parameter estimation for the Hainan MRR experiment with release point at the center of the village. About 14.8\% of the returned 4-tuples were considered outliers and are not plotted. Middle panel: Parameter estimation for the Hainan MRR experiment with release point at the periphery of the village (percentage of not-plotted outliers: 14.2\%). Right panel: Parameter estimation for the Cairns MRR experiment (percentage of not-plotted outliers: 29.4\%)}
    \label{fig:PSO-Hainan-Cairns-Parameters-4-nRuns-500-nPart-36-nGens-12-par}
\end{figure}

Figure \ref{fig:PSO-Hainan-Cairns-Parameters-4-nRuns-500-nPart-36-nGens-12-par} displays the results of the RDA-PSO method (with $N_p = 36$ and $N_g = 12$) applied to the Hainan and Cairns data shown in Tables \ref{tab:capture-table-temporal} and \ref{tab:capture-table-spatial}. The search for the minimum was performed in the following regions of the parameter space. 
\begin{align*}
&\text{(a) Hainan Center: }  {\tt k} \in [15, 55],\: {\tt q} \in [5, 30],\: {\tt p} \in [0, 15],\: s_e \in [0.001, 1];\\
&\text{(b) Hainan Edge: } {\tt k} \in [55, 95],\: {\tt q} \in [5, 40],\: {\tt p} \in [0, 5],\: s_e \in [0.001, 1];\\
&\text{(c) Cairns: } {\tt k} \in [75, 175],\: {\tt q} \in [5, 40],\: {\tt p} \in [0, 45],\: s_e \in [0.001, 1].
\end{align*}
Each panel of Figure \ref{fig:PSO-Hainan-Cairns-Parameters-4-nRuns-500-nPart-36-nGens-12-par} shows a scatter plot of the estimates of {\tt q}, {\tt p}, and ${\tt s}_{\tt e}$ as functions of the corresponding estimate of {\tt k}. Also shown (purple dots) are the resulting values of the error \eqref{eq:Error2}. Outliers, associated with an error larger than the mean plus one half (for the Cairns data) or one standard deviation (for the Hainan data), were removed. In the first two panels, the returned values of {\tt p} (squares) are near zero, which would justify the use of a simplified version of the RDA-PSO method in which {\tt p} is set to zero, to better understand the role of the other parameters. This is not the case for the Cairns study. As illustrated in Appendices \ref{sec:App_Hainan} and \ref{sec:App_Cairns}, the general rule is that in the absence of {\tt p}, the parameter {\tt q} adjusts itself to produce similar capture ratios. Estimates of ${\tt s}_{\tt e}$ (multiplication signs) are relatively high, above 34.62\%, 65.55\%, and 32.12\% in the first, second, and third panel respectively. Although some parameter estimates, especially for ${\tt s}_{\tt e}$, have large uncertainty, the range of returned values of {\tt k} is much narrower. For the Hainan study, estimates of {\tt k} are more dispersed for mosquitoes released at the edge of the village (note differences in the horizontal scale), which is to be expected since they first had to fly to the center of the village before being recaptured. In addition, the average value of {\tt k}, and thus the estimation of $D_{RDA}$, is higher for mosquitoes released at the periphery (middle panel of Figure \ref{fig:PSO-Hainan-Cairns-Parameters-4-nRuns-500-nPart-36-nGens-12-par}) than at the center (left panel of Figure \ref{fig:PSO-Hainan-Cairns-Parameters-4-nRuns-500-nPart-36-nGens-12-par}) of the village, which also is to be expected.

\begin{figure}[hbtp]
    \includegraphics[width=1.05\textwidth]{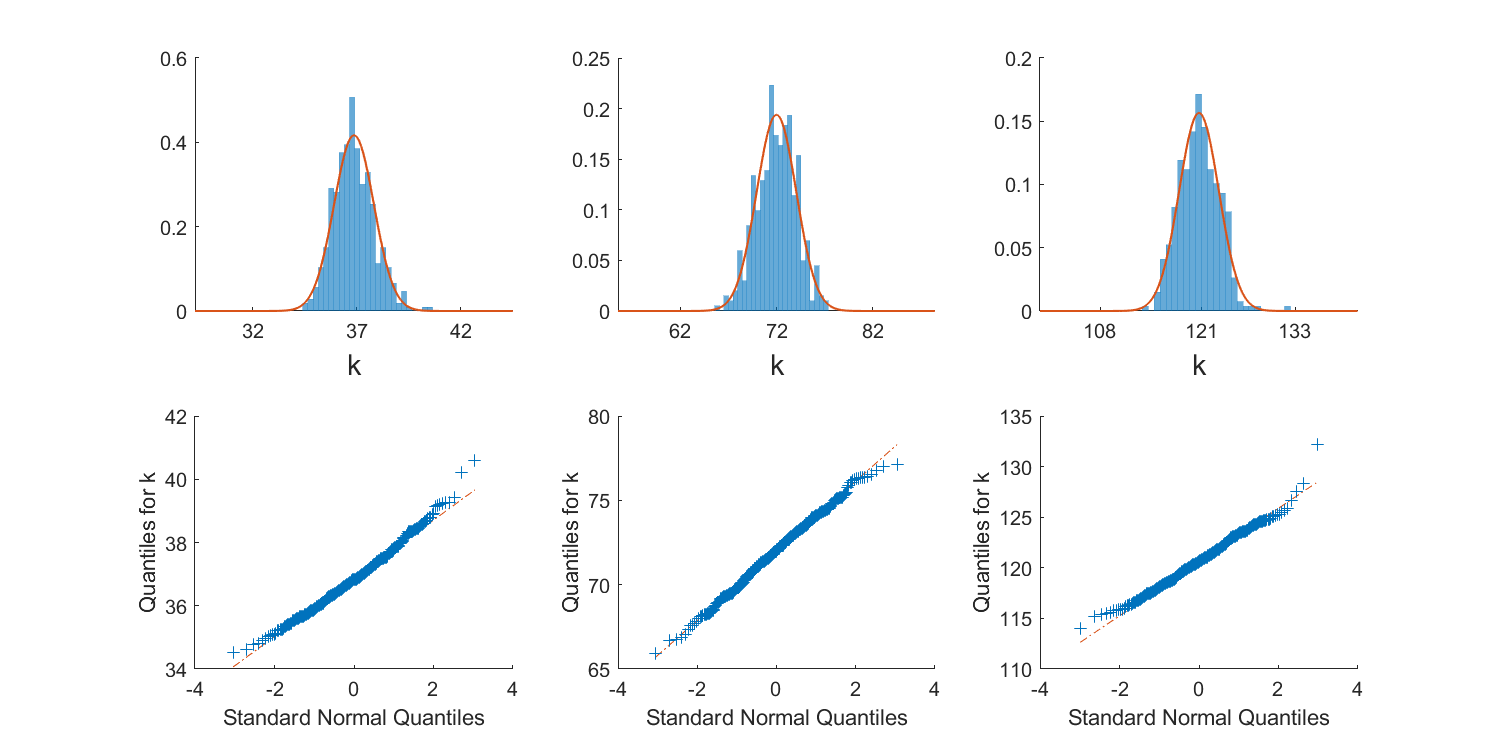}
    \caption{Empirical distributions of the {\tt k} values obtained in Figure \ref{fig:PSO-Hainan-Cairns-Parameters-4-nRuns-500-nPart-36-nGens-12-par}. The top row shows the normalized histogram of {\tt k} along with the normal distribution curve (in red) with the same mean and standard deviation as the {\tt k} data. The bottom row shows the quantiles of {\tt k} versus the theoretical quantile values from a normal distribution. Left: Hainan, center; middle: Hainan, edge; right: Cairns}
    \label{fig:PSO-All-nRuns-500-nPart-36-nGens-12-kDist}
\end{figure}

Figure \ref{fig:PSO-All-nRuns-500-nPart-36-nGens-12-kDist} shows the empirical distributions of the optimal parameter {\tt k} returned by the RDA-PSO runs, as well as associated quantile-quantile plots. These results suggest {\tt k} estimates are normally distributed. Consequently, the uncertainty on this parameter may easily be quantified by obtaining a sufficient number of estimates from the RDA-PSO method and calculating the associated mean and standard deviation. In turn, this information may be used to calculate the uncertainty on $D_{RDA}$, which is related to {\tt k} through equation (\ref{eq:DTrue_k}). 

We mentioned above that if {\tt p} = 0, the variable {\tt q} adjusts to compensate. Similarly, as {\tt s}$_{\tt e}$ decreases, larger traps are needed in order to catch the same number of mosquitoes, and vice versa. In other words, the optimal values of the parameters describing the capture sites are interconnected and consequently not individually identifiable. However, the combination of optimal parameters, {\tt q}, {\tt p}, {\tt s}$_{\tt e}$, is always associated with an optimal {\tt k} parameter that falls within the normal distribution mentioned above, leading to a reliable estimate for $D_{RDA}$. Further details are provided in Appendices \ref{sec:App_Hainan} and \ref{sec:App_Cairns} (see Figures \ref{fig:PSO-Hainan-pqCircvseff} and \ref{fig:surf-Cairns-PSO-pqCircvseff} for more information on the interconnectedness of trap-related parameters). 

\subsection{Comparison of MDT, TC, ATC, and RDA-PSO estimates} 
\label{sec:estimate_comparisons}
This section compares the diffusion coefficients estimated with the MDT, time-corrected, area-and-time-corrected, and RDA-PSO models for the above MRR experiments. All estimates are summarized in Table \ref{tab:D-fitting-corr}.

\begin{table}[hbtp]
\centering
    \caption{Values of $D$ for the Hainan and Cairns MRR experiments. Estimates of $D_{MDT}$ are obtained from Equation \eqref{eq:D_MDT} with temporal ratios given in Table \ref{tab:capture-table-temporal} and MDT values equal to 15 m, 40 m, and 77.7 m respectively. $D_{TC}$ and $D_{ATC}$ are calculated by fitting reported and corrected data, respectively, into the time-corrected and area-and-time-corrected models (\ref{eq:Q_tot_model}). Due to the limited information provided in \citep{TTW01}, the temporal ratios used to estimate $D_{ATC}$ in the Hainan studies are the same as those used to estimate the corresponding $D_{TC}$. The diffusion coefficient $D_{RDA}$ is obtained from the average of the {\tt k} values shown in Figure \ref{fig:PSO-Hainan-Cairns-Parameters-4-nRuns-500-nPart-36-nGens-12-par}, equal to 36.87, 71.99, 120.67 respectively.}
    \label{tab:D-fitting-corr}
    \begin{tabular}{lcccc}
        \toprule%
          & $D_{MDT}$ (m$^2/$day) &$D_{TC}$ (m$^2/$day) & $D_{ATC}$ (m$^2/$day)&  $D_{RDA}$ (m$^2/$day) \\
        \midrule
        Hainan - center & 32.10 & 12.81 & 36.89 & 40.78 \\
        Hainan - edge  & 201.87 & 109.36 & 119.4 & 155.48\\
        Cairns & 243.79 & 190.31 & 256.65  & 436.84 \\
        \botrule%
    \end{tabular}
\end{table}

Fitting the temporal and spatial capture ratios reported in each of Tables \ref{tab:capture-table-temporal} and \ref{tab:capture-table-spatial} into the time-corrected model (\ref{eq:Q_tot_model}) leads to estimates of the diffusion coefficient shown in the second column ($D_{TC}$) of Table \ref{tab:D-fitting-corr}. Since capture sites are not uniformly distributed across collection zones, we also show in column 3 ($D_{ATC}$) of Table \ref{tab:D-fitting-corr} the area-and-time-corrected estimates of $D$. We note however that, contrary to the Cairns study, there is not sufficient information in the Hainan data \citep{TTW01} to correct the temporal ratios after having applied the correction factor to the spatial ratios. In other words, recapture numbers are reported per zone and per day in \citep{RWW05} but not in \citep{TTW01}. For comparison, the first column lists estimates of $D$ based on averages of reported values of the MDT: between 10 and 20 m for females released at the center of the village and close to 40 m for those released at the periphery in Hainan \citep{TTW01}; between 77.6 and 77.8 m in Cairns \citep{RWW05}. The last column shows estimates of $D$ obtained by applying the RDA-PSO method.

Whereas $D_{RDA}$ and $D_{ATC}$, and separately $D_{RDA}$ and $D_{MDT}$, are reasonably close to one another (with a relative difference less than 30\%) for Hainan, there is a big difference in the two estimates for Cairns. In the latter case, as well as for the Hainan center experiment, $D_{MDT}$ and $D_{ATC}$ provide similar estimates. In addition, $D_{ATC}$ always under-estimates $D_{RDA}$. In all cases, $D_{TC}$ is much smaller than the other estimates. The latter illustrates how not compensating for the lack of spatial uniformity in the distribution of the capture sites affects estimates of mosquito dispersal. 
  
A simple intuitive interpretation of the differences in the estimates of the diffusion coefficient can be obtained using the first equation in (\ref{eq:MDT_MSDT}), which approximates the average diffusive displacement per day. For example, in the Hainan center experiment (see Table \ref{tab:D-fitting-corr}), the time-corrected estimate implies an average displacement of 6.34 m/day, whereas the RDA estimate results in a displacement of 11.32 m/day. For the Cairns experiment, the MDT estimate corresponds to a displacement of 27.67 m/day, while the RDA estimate amounts to a displacement of 37.05 m/day. Although most of these estimates are of the same order of magnitude, and are consistent with the typical flight range of \textit{Ae. aegypti} ($\lesssim 30$ m/day \citep{LMA14}), there are notable differences in their specific values.

\section{Discussion}
\label{sec:Discussion} 

This article introduces the RDA-PSO method, which is a novel computational approach that efficiently estimates the diffusion coefficient $D$ associated with mark-release-recapture (MRR) experiments. The only assumptions we make are that the dispersal of insects in those experiments is diffusive (with bias in the attracting region of the capture sites), and that MRR experiments are conducted in the absence of advection, in an environment where most of the geographic heterogeneity is associated with the presence of capture sites (houses or traps). Input data consist of temporal and spatial recapture ratios (defined in Section \ref{sec:AppRecapPerc}) and the output is an estimate of $D$, called $D_{RDA}$. The method itself (see Section \ref{sec:CompModel} as well as Appendices \ref{sec:App_ModelPhases} and \ref{sec:App_PSO}) involves a forward problem in which independent walkers diffuse from a release point and may be recaptured at sites whose locations match those used in the MRR experiment. Spatial and temporal recapture ratios obtained for each set of random walk ({\tt k}) and trap ({\tt q}, {\tt p}, and ${\tt s}_{\tt e}$) parameters are compared to the observed MRR ratios to define an error, which is minimized in the parameter space by a particle swarm optimization (PSO) method (Section \ref{sec:PSO} and Appendix \ref{sec:App_PSO}). The result is an estimate of the values of {\tt k}, {\tt q}, {\tt p}, and ${\tt s}_{\tt e}$ that best match the observed spatial and temporal recapture ratios. Our numerical explorations show that although the parameters {\tt q}, {\tt p}, and ${\tt s}_{\tt e}$ are not individually identifiable, the parameter {\tt k} is. The latter is then converted into an estimate of the diffusion coefficient $D$. 

We also introduced two other optimization approaches to estimate $D$, based on theoretical considerations: the time-corrected model (Section \ref{sec:TCM}), valid when recapture sites are uniformly distributed in the study area, and the area-and-time-corrected model (Section \ref{sec:ATCM}), useful for unevenly distributed capture sites. We explored how these estimates vary with trap location, trap density, and frequency of insect collection. More importantly, we used the time-corrected and area-and-time-corrected models as baselines against which we compared the new computational method introduced here. A third way to estimate $D$, based on the empirical MDT (Section \ref{sec:MDT}), provides results that are sometimes larger and other times smaller than those of the area-and-time-corrected model.For mosquitoes like \textit{Aedes aegypti}, which may transmit life-altering diseases such as dengue, Zika, and chikungunya, underestimating how far they may disperse once infected may have serious public health consequences.

The four methods introduced here rely on spatial and temporal ratios (defined in Equations \eqref{eq:time_pct} and \eqref{eq:sp_pct}). The temporal ratios $\tau_i$ may be viewed as weights that correct the formulas obtained from the continuous diffusion assumption in the MDT, TC, and ATC models (see Equations \eqref{eq:MDT_tot} and \eqref{eq:Q_tot_model}). By definition, these weights are proportional to the daily number of recaptured mosquitoes, which reflects pool reduction due to previous recapture, death, or dispersal beyond the area of interest. Matching the $\tau_i$'s during the optimization process provides a simple way to factor in such depletion, without having to make assumptions regarding its specific causes.

Through systematic testing, we observed that, when applied to synthetic data, $D_{RDA}$ is a very accurate estimate of the actual diffusion coefficient used to generate the data. We believe this is due to the existence of a unique minimizer in the explored parameter area, whose location is first estimated through a grid search and then refined by the PSO method (see Appendix \ref{sec:App_synth} for an example in the $({\tt k}, {\tt q})$ plane). Similar performance was not observed for $D_{TC}$, which required an increase in collection frequency to become accurate, even for uniformly distributed traps (Appendix \ref{sec:DTCM_sen} and Table \ref{tab:SyntheticD}). In addition, we found that, on average $D_{ATC}$ tends to underestimate $D$, although only correcting spatial ratios led on average to overestimates, due to round-off errors associated with the application of the correction factor (Appendix \ref{sec:DATCM_sen}). Similar results were observed with real-world data (Section \ref{sec:estimate_comparisons}), which typically involves unevenly distributed capture sites:  $D_{ATC}$ tends to be lower than $D_{RDA}$. Separately, $D_{MDT}$ tends to be lower than $D_{ATC}$ (Appendix \ref{sec:DATCM_sen} and Section \ref{sec:estimate_comparisons}), except for the Hainan - edge study (Table \ref{tab:D-fitting-corr}). In general, correcting the temporal and spatial ratios to account for uneven trap density produces values of $D_{MDT}$ and $D_{ATC}$ that are lower than partially corrected estimates that only include corrections on spatial ratios. Correcting temporal ratios thus improves $D_{ATC}$ estimates (they get closer to the known value of $D$) but worsens $D_{MDT}$ estimates. Depending on experimental design, the data collected in MRR experiments may not provide sufficient information to correct both temporal and spatial ratios. This is for instance the case for the Hainan study. 

Such a difference in performance highlights an advantage of the computational approach: the RDA-PSO method works regardless of the sampling frequency (since it calculates the ratios accordingly in the forward model), it produces estimates of $D$ that are robust to changes in the density and location of capture sites, and it does not require adjusting the observed recapture ratios, even if capture sites are unevenly distributed in the study area. In other words, the intrinsic discrete nature of RDA-PSO presents a distinct advantage over theoretical methods such as the MDT, TC, and ATC models, which are based on continuous sampling assumptions and are thus limited by collection frequency and density of capture sites.

The RDA-PSO method is stochastic in nature, leading to variability in the optimal parameter values that it returns.  This is compounded by the use of a relatively small number of swarming particles and generations to reduce computational cost. However, reliable estimates of {\tt k} may easily be obtained via repeated application of the RDA-PSO method (as seen in Figure \ref{fig:PSO-Hainan-Cairns-Parameters-4-nRuns-500-nPart-36-nGens-12-par}) and appear to be normally distributed in the parameter space (Figure \ref{fig:PSO-All-nRuns-500-nPart-36-nGens-12-kDist}). This in turn makes it relatively easy to approximate a confidence interval for $D_{RDA}$, based on different applications of the computational method. We therefore recommend running multiple (e.g. 10 or 20) simulations and removing outliers based on the returned error estimate, as discussed in Section \ref{sec:main_param_estimation}. Even though variability may be observed in the returned values of {\tt q}, {\tt p}, and ${\tt s}_{\tt e}$, we reiterate that associated values of {\tt k}, and thus the estimates of $D_{RDA}$, are not noticeably affected by the trap parameters, including the trap efficiency.

The computational cost of the method is discussed in Appendix \ref{app:computational_cost}. Although the results of Figure 1 required High Performance Computing (HPC) to demonstrate that the estimates of $D_{RDA}$ were normally distributed, a reliable value of the diffusion coefficient may be obtained in less than an hour by running a dozen simulations on a laptop.

We therefore believe the RDA-PSO method presented here provides a novel and effective way to analyze MRR data and subsequently determine insects dispersal properties in the corresponding environment. As reviewed in Section \ref{sec:dispersal}, a diffusive approximation to describe the spread of insects is appropriate and commonly used in partial differential equation (PDE) compartmental models describing the growth and spread of mosquitoes. The present method may be used not only to inform which range of values of $D$ should be included in PDE models, but also to estimate the lifetime MDT of mosquitoes like \textit{Aedes aegypti} in different environments, together with the associated risk of disease spread by these vectors. A systematic analysis of the MRR literature and estimation of associated dispersal parameters based on the RDA-PSO method introduced here is left for future work.

The above discussion situates the RDA-PSO method as a useful tool that can be incorporated into a larger array of predictive models for vector control strategies. Our current work can indeed be extended in several directions. It is straightforward to adapt our method to quantify other types of biological dispersal, as long as the process is diffusive and MRR or similar data are available. Furthermore, our code can be modified to incorporate local features such as constant directional biases (due for instance to the presence of wind or corridors such as streets in a city), attracting spots that are not capture sites and have different levels of attractiveness, etc. Although the computational cost may increase, this could pave the way to building at-scale simulations of mosquito infestations in the built environment. Weather data may be used to model the life cycle of each mosquito (see \citep{MC10} for \textit{Cx. quinquefasciatus}, \citep{BYLC15} for \textit{Cx. pipiens and Cx. tarsalis}, and \citep{LBB17} for \textit{Ae. aegypti}); the tools of Appendix \ref{sec:App_ModelPhases} can be used to digitize local maps showing buildings, trees, gardens, etc; and the forward model provides a way to simulate adult mosquitoes moving in this environment, with females being attracted to suitable oviposition sites when gravid.

\backmatter

\bmhead{Acknowledgements}

This material is partly based upon High Performance Computing (HPC) resources supported by the University of Arizona TRIF, UITS, and Research, Innovation, and Impact (RII) and maintained by the UArizona Research Technologies department. Research reported in this publication was supported in part by the University of Arizona’s BIO5 Institute Team Scholars Program. Mount Holyoke College undergraduate students, Chuhan Wang `24 and Karry Wang `23, worked on an image processing project in Summer 2022, which helped with the digitization of maps used in this work. The authors would like to thank Heidi Brown for helpful conversations on mosquito behavior. The authors would also like to thank the referees of the manuscript for their careful reading and helpful comments.

\bmhead{Data Availability}
The data that support the findings of this study are openly available in GitHub at \url{https://github.com/lidiamrad/RDA-PSO-data}. The source code is platform independent and written in Matlab.

\section*{Declarations}

\bmhead{Conflict of interest} The authors declare that they have no conflict of interest.

\newpage
\begin{appendices}
\section{Diffusion Equation}
\label{SP:diffusion}
As described in Section \ref{sec:dispersal}, diffusive mosquito dispersal means that at the microscopic level, the two-dimensional (ground) projection of the flight path of a moving mosquito consists of an alternation of straight flight segments of fixed length and angular reorientations uniformly distributed in $[0, 2\pi]$. At the macroscopic level, this corresponds to planar diffusion, described by the partial differential equation
\begin{equation}
\label{eq:diff}
\frac{\partial u}{\partial t} = D  \left(\frac{\partial^2}{\partial x^2} + \frac{\partial^2}{\partial y^2}\right) u,
\end{equation}
where $D$ is the diffusion coefficient. The associated Green's function is
\begin{equation}
u(r,t)=\frac{1}{4 \pi D t} \exp\left(- \frac{r^2}{4 D t}\right), \quad r= \sqrt{x^2+y^2}.
\label{eq:GF}
\end{equation}
Integrating expression \eqref{eq:GF} in the angular direction in polar coordinates leads to Equation \eqref{eq:gaussian_pdf_r} of the main text, describing the probability density function for the presence of a mosquito at distance $r$ from the release point, at time $t$,
\[
P_t (r)=\frac{r}{2 D t} \exp\left(- \frac{r^2}{4 D t}\right).
\]

To connect the above result to an experiment in which insects are released from a central point, we note that $u(r,t)$ may be viewed as an approximate solution of the planar diffusion equation \eqref{eq:diff} with an initial condition of the form 
\[
u(r,0)= \frac{p}{\pi} \exp\left(- p\, r^2 \right), \quad r= \sqrt{x^2+y^2},
\]
where the parameter $p$ describes the extent of the mosquito cloud at the point of release. Indeed, the solution to this initial value problem is
\[
u(r,t)= \frac{p}{\pi (1+4 p D t)} \exp\left(-\frac{p\, r^2}{1+ 4 p D t}\right).
\]
For $D >> 1$ or $t$ large, the quantity $4 p D t$ is much larger than 1 and the dependence on the initial condition through the parameter $p$ may thus be removed, leading to 
\[
u(r,t) \simeq \frac{1}{4 \pi D t} \exp\left(-\frac{r^2}{4 D t}\right),
\]
and consequently to Equation \eqref{eq:GF}.
\section{Forward Computational Model}
\label{sec:App_ModelPhases}
The forward computational model involves three phases: landscape digitization, release and recapture of diffusive agents, and analysis of the results. Each of these is described below. The code is written in Matlab, except for the first phase of digitization, which is in Python.

\subsection{Pre-processing: Landscape digitization}

In order to take into account the layout of each MRR experiment (release point, location of houses or traps, etc), we use image processing techniques to automatically extract information from the maps or sketches published by the authors of each field study. 
\begin{enumerate}
    \item \textit{Image clean-up:} Since the tools we use are unable, at this point, to precisely differentiate between some letters and houses drawn on a map, some initial cleanup (using Paint, for example) is required. 
    
    \item \textit{Gray scaling:} RGB colors are not a priority for this kind of work and we convert color images to gray-scale. Consequently, every image is read as a matrix of numbers between 0 (black) and 255 (white), or between 0 and 1, if normalized.
    
    \item \textit{Shape-detection methods:} We use the following routines found in the {\tt OpenCV} (or {\tt cv2}) and {\tt scikit-image} (or {\tt skimage}) Python packages.
    \begin{enumerate}
        \item Hough Circles ({\tt cv2.HoughCircles}) is used to detect circles (such as release sites) in an image. It returns the coordinates of the center of each circle, along with its radius.
        
        \item Information (such as centroid and equivalent radius) related to specific regions of a map is extracted with the {\tt skimapge.measure} package. This is useful for shapes that are not geometrically simple, such as house icons.
        
        \item {\tt cv2.findContours} is helpful for detecting polygons, such as rectangles that represent houses. Approximate coordinates of the centroid and vertices of a polygon may be collected if needed. 
    \end{enumerate}

\item \textit{Scaling:} For maps that have a scale bar, we average several measurements done directly on the image (using {\tt ginput} from Matlab, for example) in order to compute actual distances.
\end{enumerate}
All the data extracted from a map or sketch (centroids, rectangle vertices, image dimensions, scale) are saved in a spreadsheet and used as input for the Matlab part of the code. Index changes between Python and Matlab (Python counts from 0, Matlab from 1) are automatically processed and a 30\% margin is added on all sides of the map. A computational grid, of height {\tt H}, width {\tt W}, and mesh size {\tt h} = 0.1 meters, is defined and the coordinates of the centroids of the objects saved in the spreadsheet are added to the grid as follows. Each {\tt h} $\times$ {\tt h} square defined by the grid is referred to as a pixel.

\begin{enumerate}
    \item \textit{Capture sites} are defined by their centroids and associated with a corresponding capture-site object, consisting of two concentric regions: walkers entering the outer region are attracted towards the inner region, where they may eventually be captured.
    \item \textit{Capture site regions:} Each capture site occupies a circular domain of radius ${\tt OR} = ({\tt p} + {\tt q}) {\tt h}$, with ${\tt p} \ge 0$ and ${\tt q} > 0$. The collection region is a concentric disk of radius ${\tt IR} = {\tt q} {\tt h} \le {\tt OR}$. The attracting region occupies the remaining space within the capture site and is therefore an annulus of inner radius {\tt IR} and outer radius {\tt OR}. The movement of a walker in these regions is described in Section \ref{sec:RRDA}. Every capture site is labeled uniquely (by a specific number attached to its pixels, differentiating between attraction and collection pixels). All pixels that are not part of a capture site are categorized as being part of a region where isotropic diffusive movement occurs.
\end{enumerate}
\subsection{Processing: Release and recapture of diffusive agents}
\label{sec:RRDA}

In the simulation, agents start walking from a single release point until they are either captured, leave the region of study, or the simulation ends. Their movement corresponds to a diffusive random walk with parameter {\tt k} until they reach a capture site, at which point they might or might not be attracted to the site and/or captured depending on the trapping efficiency.

\begin{figure}[hbtp]
    \centering
        \includegraphics[width=0.9\textwidth]{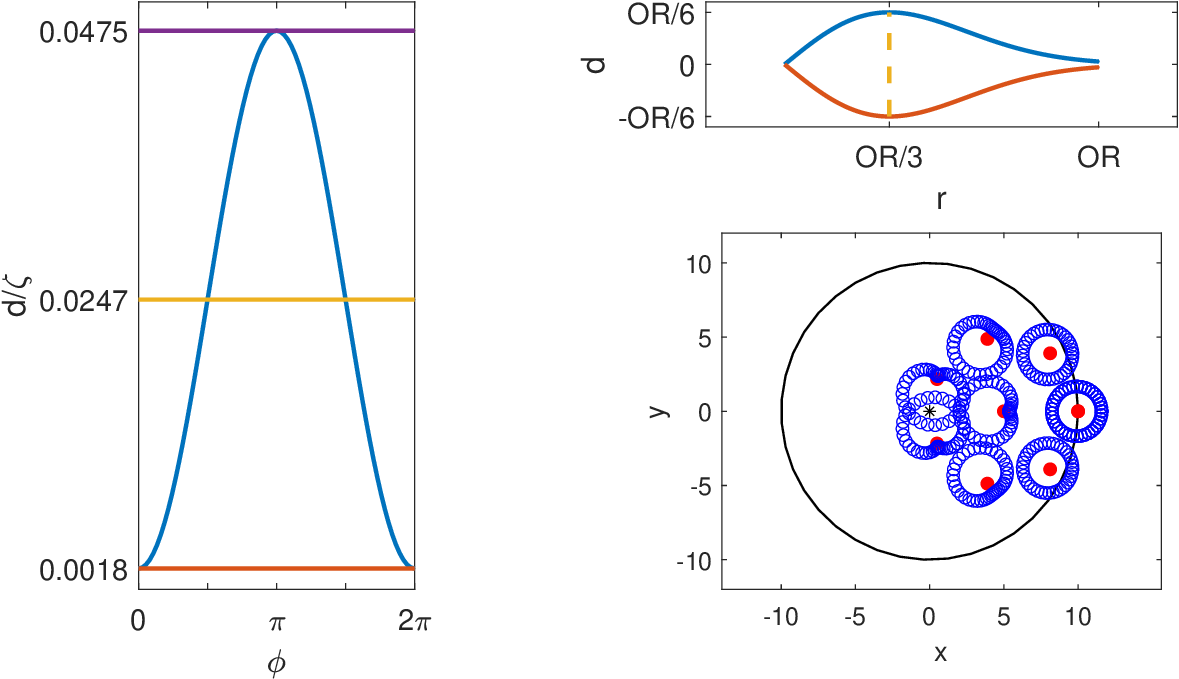}
        \caption{Left panel: Plot of {\tt d}/$\zeta$ (see Equation \eqref{eq:step_length}) as a function of $\phi$ for a walker located at $(x, y) = (2,0)$, assuming the trap center is at $(x_0, y_0) = (0,0)$. We let $\alpha = 0.2$, which leads to {\tt dBase} = 0.0247. 
        Top right panel: Range of the step sizes {\tt d} as a function of the distance $r$ between the walker location and the center of a capture site. The vertical segment at {\tt OR}/3 shows the range with largest extent. The exact value of {\tt d} depends on the angle $\phi$. Bottom right panel: Possible steps taken within a trap of outer radius {\tt OR = 10}. The red dots represent a walker's current location. The blue circles around each red dot visualize how far the walker would move if it were to head in that specific direction}
        \label{fig:distance-moved-scaled-GradF-scaledSteps-R-10}
   \end{figure}

The length of a biased step when a walker is in the attracting region of a capture site is calculated as follows. Each location within the attracting region is assigned a specific strength, associated with the gradient $\nabla F(x,y) = (S_x, S_y)$ of a bivariate normal distribution $F$ centered at the center of the capture site. Specifically, we let
\[F(x,y)  = \frac{\alpha}{\pi} e^{-\alpha[(x-x_0)^2+(y-y_0)^2]}.\]
By definition,
\[S_x = -2\alpha F(x,y) (x-x_0), \quad S_y = -2\alpha F(x,y) (y-y_0),\]
where $x_0$ and $y_0$ are the coordinates of the center of the capture site and $\alpha$ is such that $\alpha = 9/(2\, {\tt OR}^2)$, where {\tt OR} is the outer radius of the capture site. This choice of $\alpha$ guarantees that 99.7\% of the area under the graph of $F$ falls within the capture site (in other words, {\tt OR} is equal to three standard deviations of $F$), so that the effect of $F$ is felt all over the capture site.
A step from location $(x,y)$ within the attracting region is chosen to have length 
\begin{align}
{\tt  d} & = \zeta \left( {\tt dBase}+(S_x, S_y) \cdot (\cos\phi,\sin\phi) \right) \nonumber \\
& = \zeta \left( 1.01 \max_{\Omega_c} ||\nabla F||+ S_x \cos\phi + S_y\sin\phi \right), \label{eq:step_length}
\end{align}
where $\zeta$ is a positive constant that will be determined below, $\phi$ is the angle of motion of the walker measured from the $x$-axis, and $\Omega_c$ represents all of the locations inside any capture site in the simulation. The quantity ${\tt dBase}= 1.01\max_{\Omega_c} ||\nabla F||$ is chosen to be slightly bigger than $\max_{\Omega_c} ||\nabla F||$, so that the walker always changes position, that is, {\tt d} $> 0$. A simple calculation shows that $||\nabla F||$ is maximal on a circle of radius $r = 1/\sqrt{2\alpha}$ centered on each capture site, and that
\[\max_{\Omega_c} ||\nabla F(x,y)||=\frac{\sqrt{2}}{\pi}\alpha^{3/2}e^{-1/2}.\]
The expression for {\tt  d} is smaller when the walker is heading away from the center of the capture site, and larger if it is heading towards it. A plot of {\tt  d}$/ \zeta$ as a function of $\phi$ for a walker located at $(2,0)$ is shown in the left panel of Figure \ref{fig:distance-moved-scaled-GradF-scaledSteps-R-10}. Since the center of the capture site is at $(0,0)$, an angle $\phi = \pi$ indicates the walker moves towards the center of the capture site while $\phi = 0$ or $\phi = 2\pi$ indicates it is moving away from the capture region.

The length of a step taken by a walker within a capture site needs to be small compared to the size of the site, so that the walker does not immediately leave the site upon entering it. To this end, we choose $\zeta$ so that $\zeta (2.01  \max_{\Omega_c} ||\nabla F(x,y)||) = {\tt OR}/3$, which guarantees that ${\tt d} \leq {\tt OR}/3$. This gives, after substituting $\alpha$ by its chosen value in terms of {\tt OR} from above,
\[\zeta = \frac{2\pi e^{1/2}{\tt OR}^4}{2.01 \cdot 3^4}.\]
Figure \ref{fig:distance-moved-scaled-GradF-scaledSteps-R-10} (top right panel) shows the range of values of {\tt d} as a function of the distance $r$ between the location of the walker and the center of the capture site. The maximum step size occurs when $r = {\tt OR}/3$. Figure \ref{fig:distance-moved-scaled-GradF-scaledSteps-R-10} (bottom right panel) shows the step (length and direction) the walker would take from different locations within the capture site.

The code keeps track of agents that remain within the computational grid, with the final location of a captured walker considered to be the center of the capture site in which it was collected. To aid in collecting capture data, every agent ``registers'' with a capture site when it reaches its attracting region (and unregisters if it is not captured). Trap efficiency is taken into account as follows: once an agent reaches a capture site, there is a fixed chance (equal to the trap efficiency) of it sensing the trap (if the agent is in the attracting region) or being caught in it (if the agent is in the capture region). If the agent senses the trap, the next step it takes is biased towards its center, as described above. If not, it takes a normal step of length {\tt k}$\cdot${\tt h}. This process continues until the agent is either captured or leaves the capture site. Capture data can be saved by hour or by day, whereas movement is always tracked by hour. It is worth noting that each walker has no direct interaction with other walkers. Capture ratios are averaged over a few simulations, typically 5, to reduce intrinsic noise.

\subsection{Post-processing: results analysis}
In the post-processing phase, the data saved during the random walk simulations are analyzed and compared to the results from field experiments. The type of data saved depends on the experiments we aim to simulate. For example, recapture numbers are saved in arrays whose rows are zone labels and columns are time. These are then compared to observed data by estimating the error $E$ defined in \eqref{eq:Error2}. 

\section{Particle Swarm Optimization}
\label{sec:App_PSO}
The PSO computational method is originally attributed to Kennedy and Eberhart \citep{KE95, SE98}, and was developed to simulate social behavior (hence the `swarm' wording). We chose this approach because the algorithm can search large parameter spaces without making assumptions on the problem being optimized. In particular, there are no differentiability requirements, since the method is not gradient-based. Although the procedure itself does not guarantee an optimal solution is found, we always find a minimizer in our applications, using only a small number of particles and a few generations. In addition, this optimization approach typically does not get stuck at a local minimum.

\subsection{Description of the method}
The particles move around the search space according to a simple formula. These movements are informed by each particle's best estimate of the minimizer location, $\mathbf{x}_{localMin}$, as well as by the best estimate of the entire swarm, $\mathbf{x}_{globalMin}$, where $\mathbf x$ refers to the vector of coordinates $({\tt k},{\tt q},{\tt p},{\tt s}_{\tt e})$. Here, ``best estimate'' corresponds to the smallest error $E$. More specifically, $\mathbf{x}_{localMin}$ is the position in parameter space where the error found by the particle itself is the smallest so far, whereas $\mathbf{x}_{globalMin}$ is the position where the error evaluated so far by all of the particles is smallest.

The 4-dimensional position $\mathbf{x}$ of each particle is updated according to $\mathbf{x}_{new} = \mathbf{x}_{old} + \mathbf{v}_{new}$, where
\[\mathbf{v}_{new} = w \,\mathbf{v}_{old} + \phi_p  \mathbf{r}_1  *(\mathbf{x}_{localMin} - \mathbf{x}_{old}) + \phi_g  \mathbf{r}_2  *(\mathbf{x}_{globalMin} - \mathbf{x}_{old}),\]
and $\mathbf{r}_1$ and $\mathbf{r}_2$ are 4-dimensional vectors of independent random numbers drawn from the uniform distribution in the interval $[0,1]^4$. Note that ``$*$" here denotes element-wise multiplication.
The parameters are $w$, the ``inertia" weight, $\phi_p$, the ``cognitive" coefficient, and $\phi_g$, the ``social" coefficient. These heuristic coefficients are chosen so that $0<w<1$ and $\phi_p, \phi_g \geq 1$, reflecting a balance between moving towards the ``best seen" position and swarming for a more effective search. In particular, with $w<1$, the first term acts to slow down the particle when getting close to a minimum. The remaining two terms move it generally in the direction of its own local best position (second term) and the swarm's global best position (third term), with an element of randomness included in that motion. We initiate $\mathbf{v}$ randomly in $[-1, 1]^4$, and denote the resulting vector from the previous step by $\mathbf{v}_{old}$. For all of the simulations presented here, we set $w = 0.4$ and $\phi_p = \phi_g = 1.2$. A new generation of particles is obtained after all of the positions of the previous generation have been updated once according to the above formula.

\subsection{Computational cost}
\label{app:computational_cost}
The code is not yet optimized for usability, but we believe its current computational cost is reasonable. Table \ref{tab:CompCost} reports on the time elapsed (using the MATLAB {\tt tic} and {\tt toc} commands) to obtain an estimate of {\tt k} from several runs of the RDA-PSO method. We report the average of {\tt k} over multiple runs (performed in parallel to save computational time), as well as an interval corresponding to one standard deviation about the mean value of the {\tt k} estimates. Simulations were run for the Hainan - center field experiment (left panel of figure \ref{fig:PSO-Hainan-Cairns-Parameters-4-nRuns-500-nPart-36-nGens-12-par}). As usual, the parameter tuples associated with an error that was one standard deviation above the mean error were considered outliers and removed. Different configurations (laptop or high-performance computer (HPC), 2- or 4-parameter optimizations) were analyzed. Since the optimal value of {\tt p} obtained for this study is close to zero, the 2-parameter and 4-parameter runs are expected to produce estimates of {\tt k} that are close to one another, even though this is not generally the case.

\begin{table}[hbtp]
\centering
    \caption{Elapsed time for 4-parameter ({\tt k}, {\tt q}, {\tt p}, ${\tt s}_{\tt e}$) and 2-parameter ({\tt k}, {\tt q}) optimizations (with {\tt p}=0 and ${\tt s}_{\tt e}$=1 for the latter), on a laptop with 8 workers and on an HPC using 1 node with 32 workers. The mean value of {\tt k} and its range within one standard deviation are also reported.}
    \label{tab:CompCost}
    \begin{tabular}{llcccc}
        \toprule%
          & & number of  & elapsed time & $\overline{{\tt k}}$ & $[\overline{{\tt k}} - \sigma,\: \overline{{\tt k}} + \sigma]$ \\
          & & RDA-PSO runs & (in minutes) & &\\
        \midrule
        4-parameter opt.  & laptop & 1 & 10.53 & 36.99 & -- \\
          & & 10 & 38.65 & 36.93 & [36.3, 37.56]\\
        & HPC & 10 & 20.67 & 37.43 & [36.34, 38.51]\\
        & & 100& 69.93 & 36.95 & [36.02, 37.89]\\
        & & 500& 277.95 & 36.78 & [35.83, 37.73]\\
        2-parameter opt. & laptop & 10 & 33.17 & 36.46 & [35.62, 37.29]\\
        & HPC & 100 & 44.67 & 36.27 & [35.58, 36.97]\\
        \botrule%
    \end{tabular}
\end{table}

For all of the configurations considered, the average of {\tt k}  is close to the one obtained using 500 RDA-PSO runs, which is the most accurate and requires the most computational time. One RDA-PSO run with 36 particles and 12 generations corresponds to 36*13 simulations, each of which is the average of 5 independent iterations of the forward model. So one RDA-PSO run corresponds to 2,340 release-recapture experiments. Looking at the elapsed time for 1 RDA-PSO run on the laptop (without parallelization) in Table \ref{tab:CompCost}, this is equivalent to 0.27 seconds per release-recapture experiment. The computational cost of each RDA-PSO run goes down as the number runs, executed in parallel, increases. For example, for 10 runs on the HPC, each run takes about 124 seconds. For 500 runs, each one takes about 33.35 seconds.
\section{Virtual MRR experiment}
\label{sec:App_synth}
This section illustrates how the different methods introduced in this article perform on a virtual MRR experiment. To this end, we first obtain synthetic capture ratios by simulating the forward problem with fixed parameters and averaging the resulting ratios of 100 simulations. This leads to very robust ratios that fully capture the observed dispersal. Consequently, all of the models, except for TC which uses uncorrected ratios, estimate values of $D$ that are reasonably close (of the order of 10\% or less) to the known diffusion coefficient, thereby providing a validation of each of these approaches. The RDA-PSO method, iterated 100 times in this case, is highly accurate and provides the best results. Appendix \ref{SP:ATCM_prop} compares the different methods in the case of {\em synthetic} MRR experiments. By synthetic, we mean experiments that use realistic trap placements and observed ratios (obtained without averaging the forward model 100 times), but for which the diffusion coefficient is still known. 

\subsection{Virtual physical landscape}

\begin{figure}[hbtp]
\centering
    \includegraphics[width=0.85\textwidth]{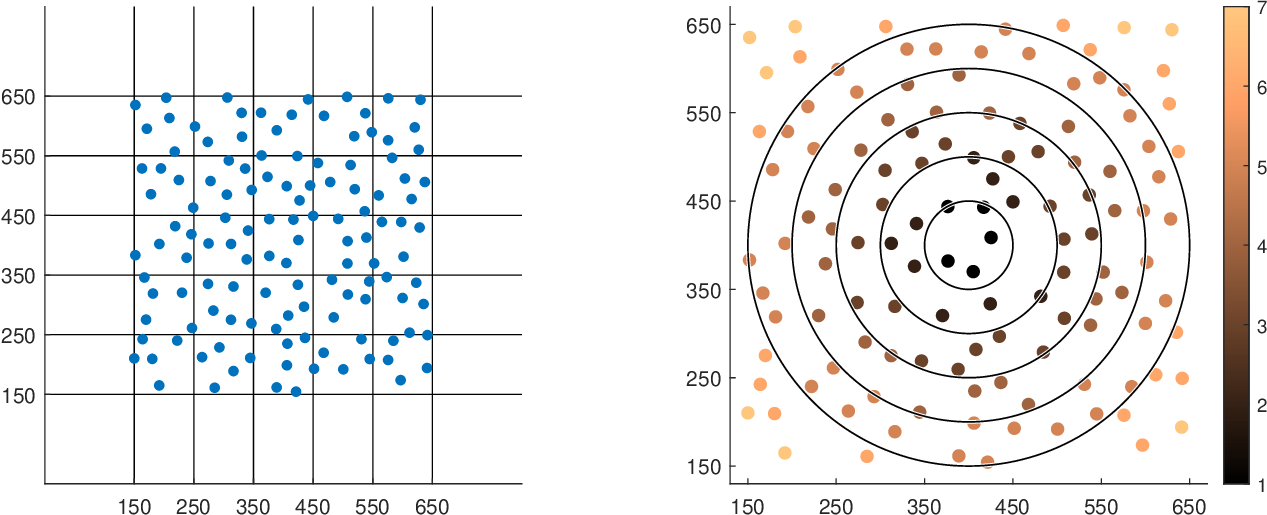}
    \caption{Left panel: Virtual map with 5 capture sites per region of size 100 m $\times$ 100 m. Margins equal to 150 m are shown on all sides. Right panel: Concentric zones of width 50 m superimposed on the capture sites shown in the left panel. In this example, there are 5, 9, 22, 25, and 39 capture sites in zones 1 through 5, respectively}
    \label{fig:traps-by-region-TrD-5}
\end{figure}

We assume the virtual MRR experiment takes place in a region of size 500 m $\times$ 500 m, which we subdivide into smaller squares of side $100$ m each. Each of these square regions contains 5 uniformly distributed capture sites (see left panel of Figure  \ref{fig:traps-by-region-TrD-5}). Concentric zones around the release site, which are used to calculate the spatial and temporal recapture ratios, are 50 m apart and visualized in the right panel of Figure \ref{fig:traps-by-region-TrD-5}. 

\subsection{Generation of MRR data for the virtual experiment}

In order to limit the effect of randomness in the generated data, we run 100 simulations of the forward problem with the above setup and use average recapture ratios as the observed ratios. In each of the 100 simulations, $N=$100,000 agents are released from the middle of the grid. Each moves independently of the others and performs a random walk with parameters {\tt k} = 100 and {\tt q} = 30 for a total duration of 20 days. Trap efficiency is set at 100\% and {\tt p} = 0. Table \ref{tab:capture-table-virtual} lists the average capture ratios, by day (with a count every other day) and by zone, resulting from these simulations. The selected value of {\tt k} corresponds to a diffusion coefficient $D = 300$ m$^2$ per day.

\begin{table}[hbtp]
    \caption{Temporal (left) and spatial (right) ratios of captured mosquitoes for the virtual experiment.}
    \label{tab:capture-table-virtual}
    \begin{tabular*}{0.7\textwidth}{@{\extracolsep\fill}c@{\extracolsep\fill}c@{\extracolsep\fill}c@{\extracolsep\fill}c}
    \toprule%
    Day &   Temporal ratios, $\tau_i$ & Zone (in m) & Spatial ratios, $\sigma_j$ \\
    \cmidrule{1-2}\cmidrule{3-4}%
    2 & 25.65\% &  $(0, 50)$ & 52.36\%\\
    4 & 20.91\% & $(50, 100)$ & 24.03\%\\
    6 & 15.35\% & $(100, 150)$ & 16.86\%\\
    8 & 11.39\% & $(150, 200)$ & 5.14\%\\
    10 & 8.43\% & $(200, 250)$ &1.61 \%\\
    12 & 6.23\%\\
    14 & 4.54\%\\
    16 & 3.32\%\\
    18 & 2.42\%\\
    20 & 1.76\%\\
    \botrule
    \end{tabular*}
\end{table}

\subsection{Application of the MDT model}
We calculate the empirical MDT by multiplying the spatial ratios $\sigma_j$ element-wise by the average distance to each corresponding zone, that is
\[\text{MDT} = \sum_{j=1}^{n_z} \sigma_j \:\frac{r_{j-1}+r_j}{2}.\]
Note that in the above formula, the distance at which each mosquito is recaptured is approximated by the mean radius of the zone in which it was caught. The resulting value of $D_{MDT}$ is approximately $278.3$ m$^2$ per day, which is an under-estimate. Using corrected spatial and temporal ratios leads to $D_{MDT} \simeq 320.3$ m$^2$ per day, which is an over-estimate.

\subsection{Application of the time- and area-and-time-corrected models}
Fitting the data obtained for the temporal and spatial capture ratios into the time-corrected model (\ref{eq:Q_tot_model}) gives $D_{TC} \simeq 234.6$ m$^2$ per day, which is an under-estimate. Although the traps are uniformly distributed in each square simulation region, the number of traps in each annular zone is not exactly proportional to its surface area. Correcting for this imbalance leads to an area-and-time-corrected estimate of the diffusion coefficient, $D_{ATC} \simeq 331.4$ m$^2$ per day, which is an over-estimate. 

\begin{figure}[hbtp]
\centering
    \includegraphics[width=\textwidth]{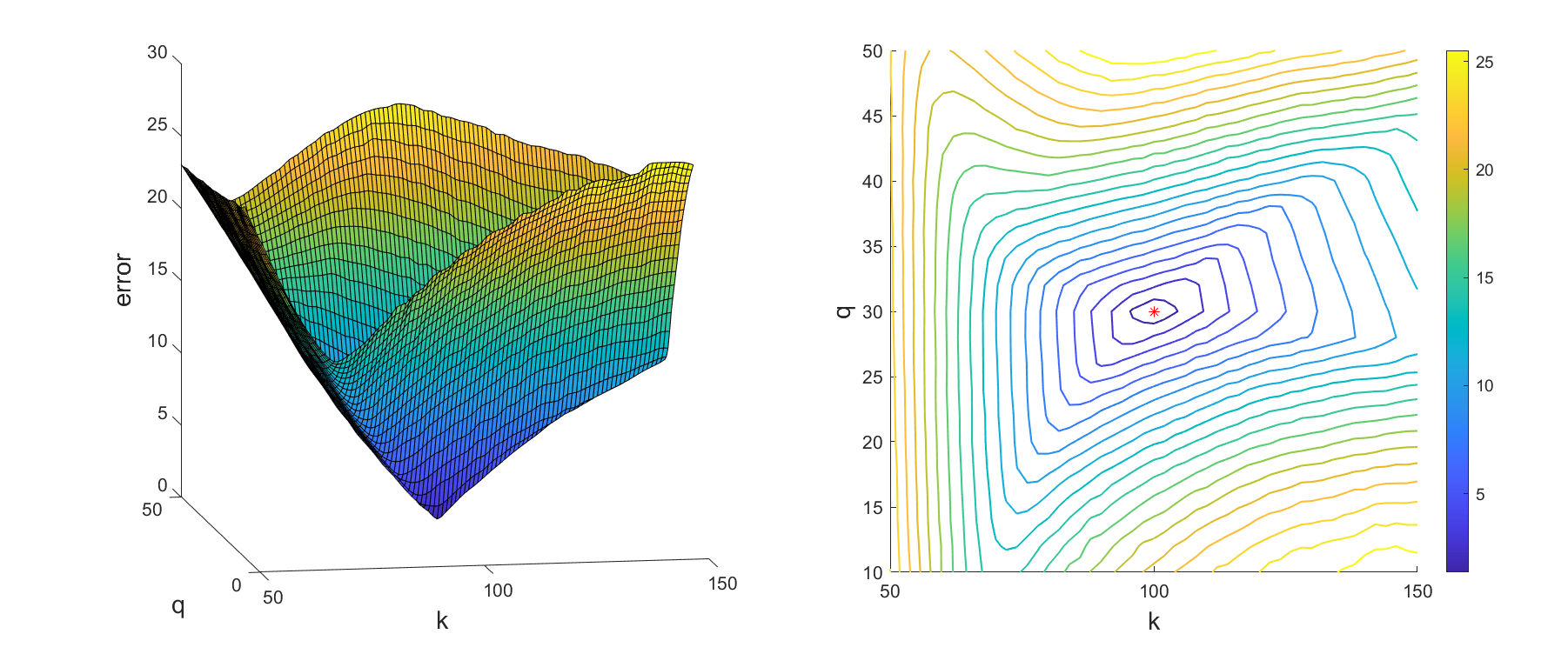}
    \caption{Left panel: Error landscape produced by the grid search for the synthetic data of Table \ref{tab:capture-table-virtual}. The error $E$ is calculated over a uniform mesh of size $51 \times 21$, with values of {\tt k} in the interval $[50, 150]$ and values of {\tt q} in the interval $[10, 50]$. Note that the surface is plotted using cubic interpolation over the actual data. Right panel: Contour plot of $E$ showing the estimated minimizer (red star) near {\tt k} = 100 and {\tt q} = 30. Here, the parameter {\tt p} is set at ${\tt p} = 0$ and the trap efficiency at ${\tt s}_{\tt e} = 100$~\%}
    \label{fig:virtual-e2}
\end{figure}

\subsection{Application of the computational method}
The RDA-PSO method involves two consecutive steps: a grid search to identify a region that contains optimal parameter values, followed by particle swarm optimization to refine the parameter estimation.

\subsubsection{Grid search}
The initial search for the minimizer of the error $E$ defined in Equation \eqref{eq:Error2} of the main text, with observed recapture ratios given in Table \ref{tab:capture-table-virtual} is performed over a $51 \times 21$ grid with values of {\tt k} in the interval $[50, 150]$ and values of {\tt q} in $[10, 50]$. For each grid point, the forward problem (100,000 agents with capture ratios averaged over 5 simulations) is run with ${\tt p}=0$ and a trap efficiency equal to 100 \%. As shown in Figure \ref{fig:virtual-e2}, the error landscape has a single minimizer (where the error is 0.09), whose location is consistent with the values of {\tt k} and {\tt q} used to create the synthetic MRR data. 

\subsubsection{Particle swarm optimization} Once the general location of the error minimizer is identified by the systematic grid search procedure, the particle swarm optimization (PSO) method is applied to pinpoint an estimate of the location of the minimizer. Here, the PSO routine is run with 15 particles over 5 generations, in the parameter space region defined by {\tt k} $\in [70, 130]$ and {\tt q} $\in [16, 44]$. Out of 100 PSO runs, we keep the results whose error is less than the mean error plus one standard deviation, thus accepting 86\% of the results. The average value of {\tt k} = 99.85 with a standard deviation of 0.76 and the average value of {\tt q} = 29.92 with a standard deviation of 0.21. The error ranges between $E = 0.0856$ and $E = 0.5562$ (after removing the outliers). These numerically inferred average parameters are within a few percents of the actual values used to create the data ({\tt k} = 100 and {\tt q} = 30). If the direct problem in the RDA-PSO method is run with N=10,000 walkers instead of 100,000, we obtain {\tt k} = $99.89 \pm 0.996$ and {\tt q} = $29.92 \pm 0.29$ with the error ranging between 0.2464 and 0.6648.

\subsubsection{Estimation of the diffusion coefficient}
Using the average value of {\tt k} found by the RDA-PSO model, {\tt k} = 99.85, leads to an estimate of the diffusion coefficient $D_{RDA} = 299.1$ m$^2$ per day, which is extremely accurate and presents a significant improvement on the $D_{MDT}$, $D_{TC}$, and $D_{ATC}$ estimates mentioned above. When implemented on synthetic data, the RDA-PSO method is thus able to estimate the diffusion coefficient with high accuracy, as shown in Table \ref{tab:D-virtual}, which summarizes all the diffusion coefficient estimates obtained in this section, along with their signed relative error.

\begin{table}[hbtp]
\centering
    \caption{Estimates of $D$ for the virtual experiment. Exact $D$ is 300 m$^2/$day.}
    \label{tab:D-virtual}
    \begin{tabular}{lcc}
        \toprule%
        & estimate & relative error \\
        & (in m$^2/$day) & from exact $D$ \\
        \midrule
          $D_{MDT}$ & 278.3 & $-7.23\%$\\
          $D_{MDT}$ (corrected) & 320.3 & $6.8\%$\\
          $D_{TC}$ & 234.6 & $-21.8\%$\\
          $D_{ATC}$ & 331.4 & $10.47\%$\\
          $D_{RDA}$ & 299.1 & $-0.3\%$\\
        \botrule%
    \end{tabular}
\end{table}

\section{Hainan study}
\label{sec:App_Hainan}
This section describes the layout of two real-world MRR experiments conducted on Hainan island in China \citep{TTW01}. Mosquitoes were released from two different sites, one at the center (where houses are concentrated) and another at the edge of a village. Both male and female mosquitoes were released and, assuming they were released in equal numbers, 3009 female mosquitoes left from the center and 3076 from the edge. Released insects were recaptured at every house in the village over a period of 6 days. 
\begin{figure}[hbtp]
\centering
    \includegraphics[width=0.75\textwidth]{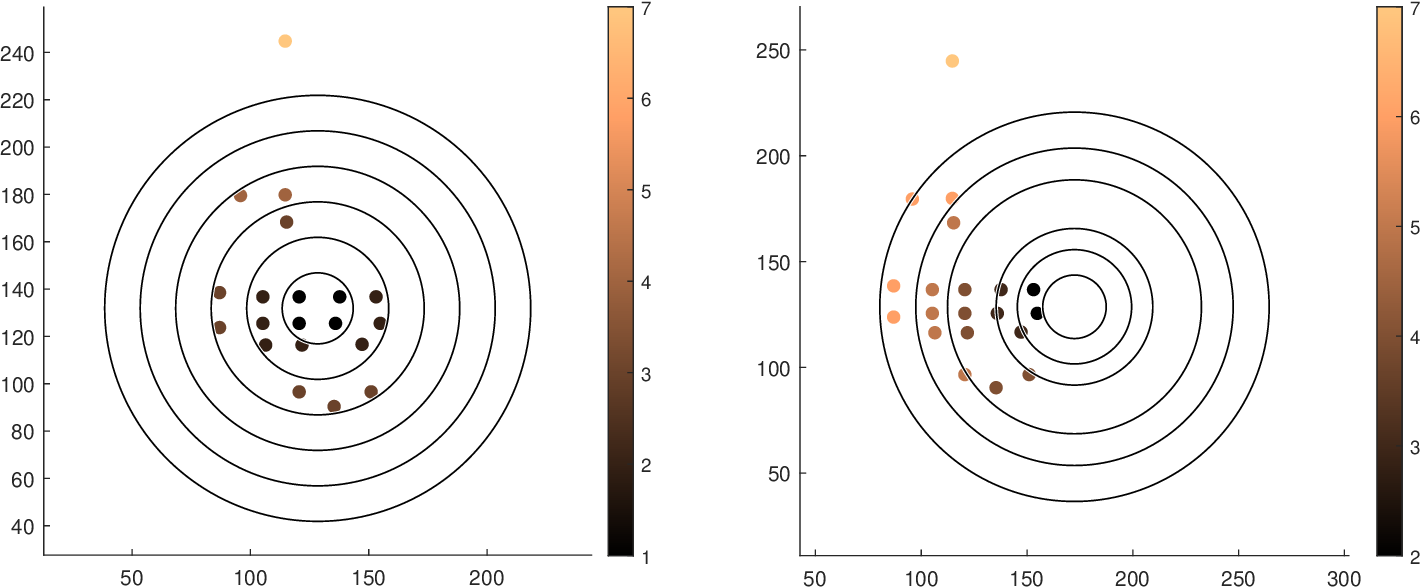}
    \caption{Left panel: Capture sites and concentric zones (15 m apart) associated with the central release site. There are 4, 7, 6, 2, 0, and 0 capture sites in zones 1 through 6 respectively. Each zone is labeled by the number indicated on the color bar on the right. Right panel: Same as the left panel but for the release site at the edge of the village. There are 0, 2, 3, 5, 5, and 4 capture sites in zones 1 through 6 respectively. Note that the zones here are not equidistant in order to match the number of houses per zone given in the original article. The concentric circles have radii $15, 27, 37, 60, 75, 92$ m respectively. In both maps, the furthest house from the release site is excluded from the $D_{ATC}$ calculations but included in the $D_{RDA}$ calculations}
    \label{fig:traps-by-region-Hainan}
\end{figure}

The map of the village provided in \citep{TTW01} was used to plot Figure \ref{fig:traps-by-region-Hainan}, in which each dot represents a house. Six concentric zones, centered either on the central (left panel of \ref{fig:traps-by-region-Hainan}) or peripheral (right panel of Figure \ref{fig:traps-by-region-Hainan}) release point are also shown. Recaptured numbers and recapture ratios of female mosquitoes released from both locations are listed in Tables \ref{tab:capture-table-temporal} and \ref{tab:capture-table-spatial} of the main text. The size of the zones associated with the release in the center of the village was adjusted to take into account that the house furthest from the release point did not catch any mosquitoes. Because the surface area of each zone significantly affects estimates of $D_{ATC}$, it is indeed important to only include capture sites that record mosquito presence. In addition, since this house is more than 15 meters away from the previous zone boundary, including it would lead to a zone of very large surface area with few capture sites, thereby leading to large and unrealistic corrections to the spatial (and temporal) ratios. Consequently, for the edge release, we also assume that this house did not catch any mosquitoes, i.e. that the 4 houses in the first 17 meters of zone 6 capture all of the mosquitoes for that zone. The furthest house is included in the RDA-PSO simulations since the surface area of the zones is not relevant for that method.

\subsection{Grid search} 
\begin{figure}[hbtp]
\centering
    \includegraphics[width=\textwidth]{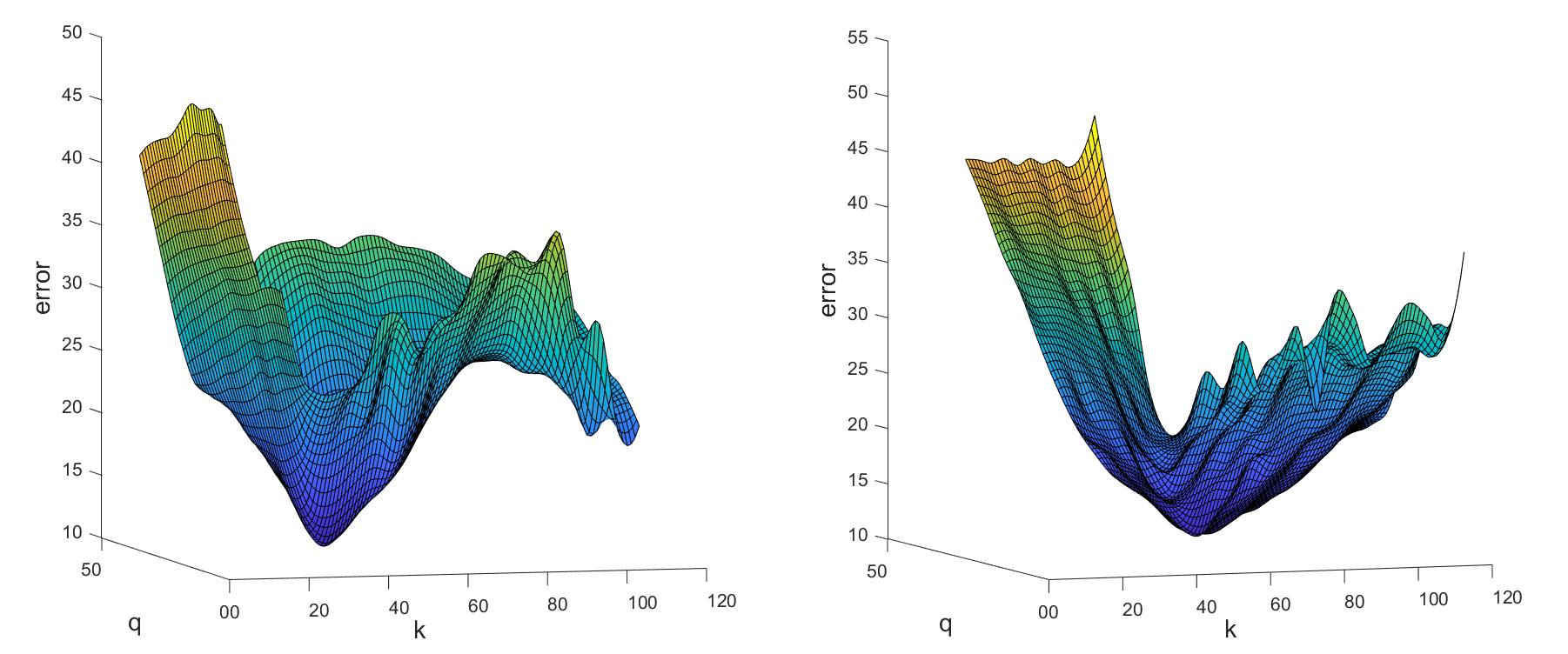}
    \caption{Error landscape calculated over a uniform mesh in the $({\tt k},{\tt q})$ space. Left panel: Error obtained for ({\tt k}, {\tt q}) in $[5,105]\times[3, 43]$ where mosquitoes were released from the center. Right panel: Error obtained for ({\tt k}, {\tt q}) in $[15,115]\times[3, 43]$ where mosquitoes were released from the edge. Note that the surface is plotted using cubic interpolation over the actual data}
    \label{fig:surf-Hainan-females}
\end{figure}

The first step of the optimization procedure consists in running simulations over a uniformly distributed sample of {\tt k} and {\tt q} while setting {\tt p} = 0 and ${\tt s}_{\tt e} = 100\%$. We take ({\tt k}, {\tt q}) in the range $[5,105]\times[3, 43]$ for the center release and $[15,115]\times[3, 43]$ for the edge release, with a total of 231 parameter pairs for each (21 values for {\tt k} and 11 values for {\tt q}). The error landscape, using equation (\ref{eq:Error2}), is shown in the left panel of Figure \ref{fig:surf-Hainan-females} for the center release and the right panel of Figure \ref{fig:surf-Hainan-females} for the edge release. This helps narrow down the PSO search to $[15, 55]\times[5, 30]$ and $[75, 95]\times[5, 40]$ respectively, with the ranges of {\tt p} calculated so that no traps overlap and {\tt s}$_{\tt e}$ ranging from 0.1\% to 100\%.

\subsection{PSO results} 
\begin{figure}[hbtp]
\centering
    \includegraphics[width=0.45\textwidth]{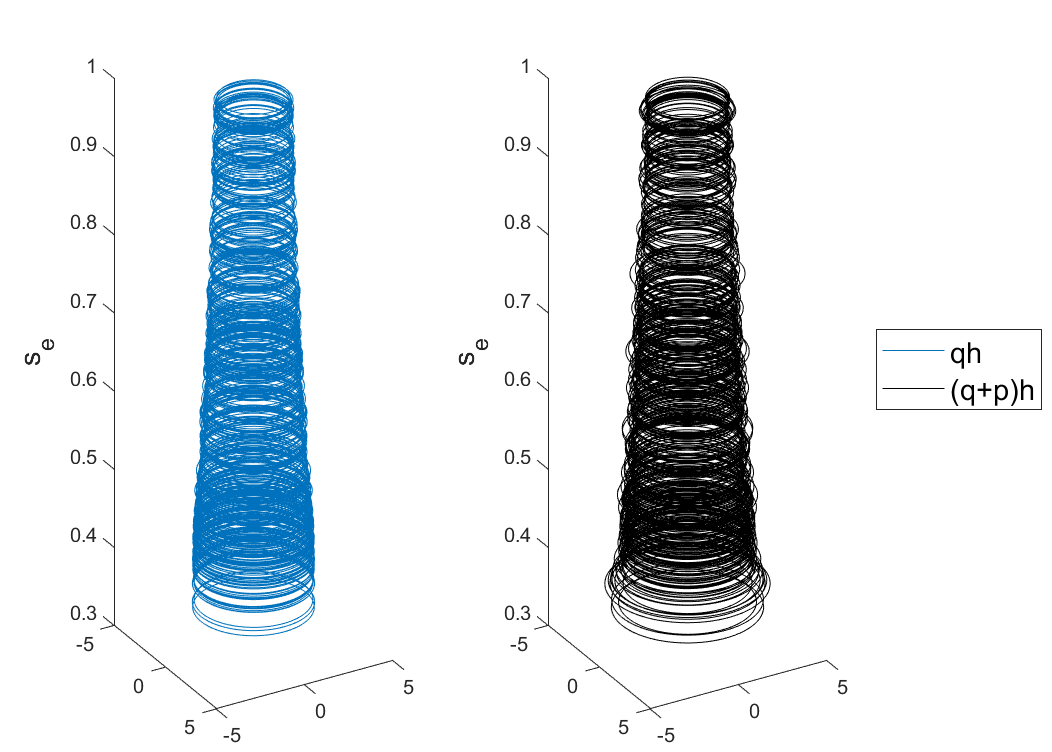}\hspace{0.1in} \includegraphics[width=0.45\textwidth]{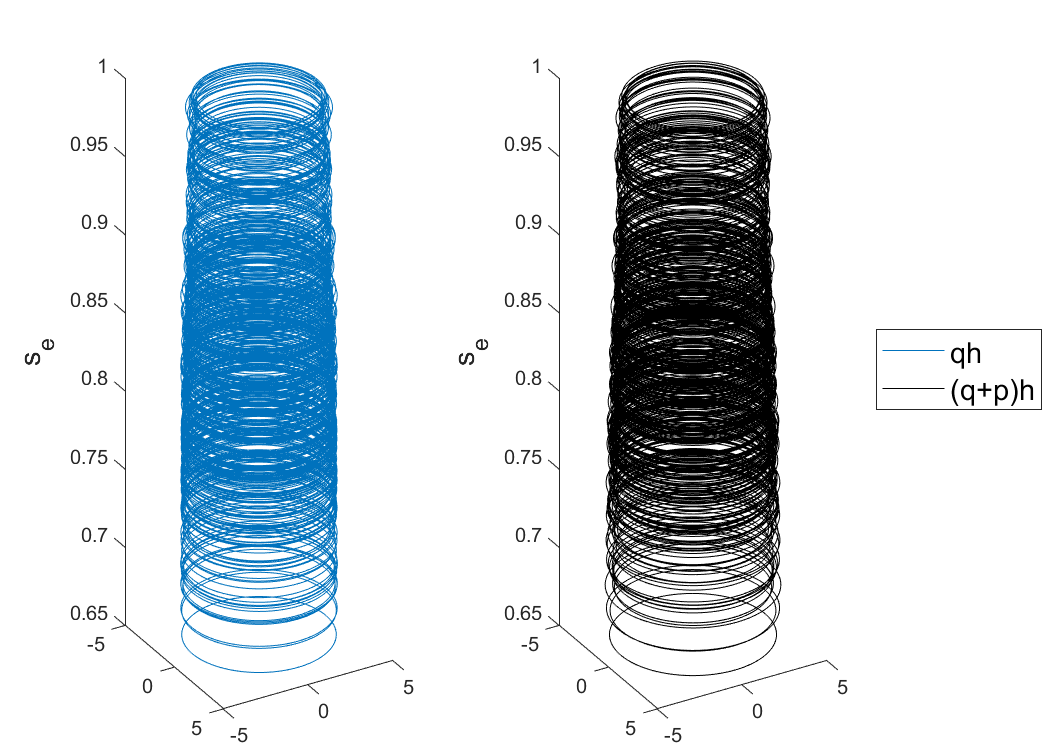}
    \caption{PSO results from 500 runs (as usual, outliers associated with a large error are removed; see main text for details). The optimal trap efficiency {\tt s}$_{\tt e}$ is shown as a function of the optimal inner radius {\tt q} $\cdot$ {\tt h} (blue circles) and of the optimal outer radius ({\tt q}+{\tt p}) $\cdot$ {\tt h} (black circles) of each capture site. Left panel: Hainan study with central release. Right panel: Hainan study with edge release}
    \label{fig:PSO-Hainan-pqCircvseff}
\end{figure}

As shown in the main text, the optimal values of the parameter {\tt k} obtained from the RDA-PSO method follow a normal distribution, so that the average of a number of runs can yield an accurate estimate of the diffusion coefficient. This is not the case for the remaining parameters, {\tt q}, {\tt p} and ${\tt s}_{\tt e}$. However, we observe that these parameters do follow a certain pattern. Specifically, when the value of ${\tt s}_{\tt e}$ is high, the {\tt q} and {\tt p} values are low, and vice versa as shown in Figure \ref{fig:PSO-Hainan-pqCircvseff}. That is, the size of the capture site adjusts according to its efficiency, so that the same number of mosquitoes are recaptured.

\section{Cairns study}
\label{sec:App_Cairns}
\begin{figure}[hbtp]
\centering
    \includegraphics[width=0.4\textwidth]{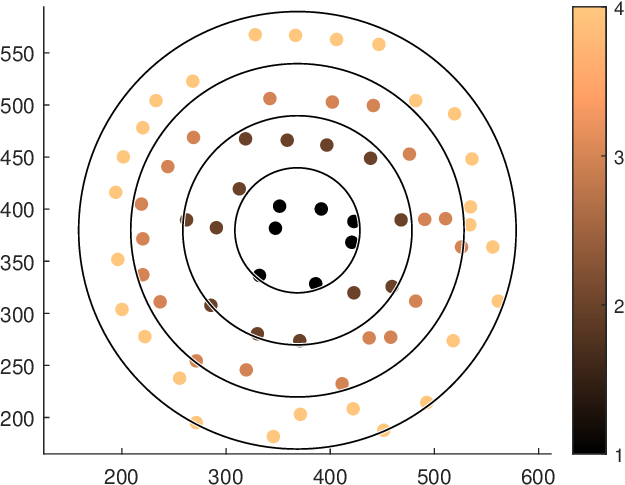}
    \caption{Capture sites and concentric zones (50 m apart) associated with the Cairns study. There are 7, 13, 19, and 27 capture sites in zones 1 through 4 respectively. Each zone is labeled by the number indicated on the color bar on the right}
    \label{fig:traps-by-region-Cairns}
\end{figure}

The zones associated to the experiment carried out in Cairns are drawn in figure 1 of \citep{RWW05}, but their boundaries are not exactly circular, probably due to an anisotropic scaling of the figure during printing. In order to keep the correct number of capture sites in each zone, the $y$-scale of the map provided in \citep{RWW05} was slightly compressed on each side of the release site. This made the zones circular, while keeping each capture site in its associated zone. The result is shown in Figure \ref{fig:traps-by-region-Cairns}; there are 7, 13, 19, and 27 capture sites in each zone respectively. The study site includes streets, which could limit mosquito spread or serve as corridors that facilitate dispersal. Because our simulations assume a homogeneous landscape away from capture sites, the estimates of $D$ we obtain may not be accurate. They can however be compared to one another, which is the purpose of the present discussion.

Collection of mosquitoes started 2 days after release, including on the release site. Collection was made on days 5, 8, 11, 15 by replacing the capture sites. A total of 67 mosquitoes were captured, 15 of them from the release site. Since these 15 mosquitoes never left the release site, they are excluded from our data. Out of the 52 caught outside the release area, 22 were caught within 50 m, 33 within 100 m (by day 8), and only 12 were caught further than 100 m over the whole period (15 days). Together with additional information from figure 2 of article \citep{RWW05}, these statements are summarized in Table \ref{tab:capture-table-Cairns}. 
\begin{table}[hbtp]
    \caption{Number of captured mosquitoes by capture day and zone, summarized from \citep{RWW05}.}
    \label{tab:capture-table-Cairns}
    \begin{tabular*}{0.7\textwidth}{@{\extracolsep\fill}c|cccc|@{\extracolsep\fill}c@{\extracolsep\fill}}
    \toprule%
    Zone (in m) $\backslash$ Day  & 5 & 8  & 11 & 15 & Number Recaptured \\
    \cmidrule{1-6}
    $(0, 50)$ & 10 & 12 & 1 & 1 & 24 \\
    $(50, 100)$ & 8 & 3 & 3 & 2 & 16 \\
    $(100, 150)$ & 1 & 3 & 2 & 1 & 7 \\
    $>150$  & 1 & 0 & 3 & 1 & 5 \\
    \cmidrule{1-6}
    Number Recaptured & 20 & 18 & 9 & 5 & 52\\
    \botrule
    \end{tabular*}
\end{table}

\subsection{Estimating the area-and-time-corrected diffusion coefficient} The area-and-time-corrected estimate of the diffusion coefficient, $D_{ATC}$, is calculated as follows. Using Table \ref{tab:capture-table-Cairns}, create a matrix $C$ containing the counts, where the rows represent recapture numbers per zone and the columns represent recapture numbers per day. Each row of $C$ is then multiplied by 
\begin{equation}
    c_j = \frac{CF_j}{nT_j} = \frac{nT_{tot}}{nT_j} \, \frac{A_j}{A_{tot}},
    \label{eq:CorrectionMatrixCount}
\end{equation}
to get the corrected count matrix $C^c$. To find the corrected temporal ratios $\tau^c_i$, we add the columns of $C^c$ and to find the corrected spatial ratios $\sigma_j^c$, we add the rows of $C^c$. These corrected ratios are then fitted in equation \eqref{eq:Q_tot_model_fit} to obtain the estimate of $D_{ATC}$ provided in Table \ref{tab:D-fitting-corr}.

\subsection{Grid search and PSO results} 
\begin{figure}[hbtp]
\centering  
    \includegraphics[width=0.42\textwidth]{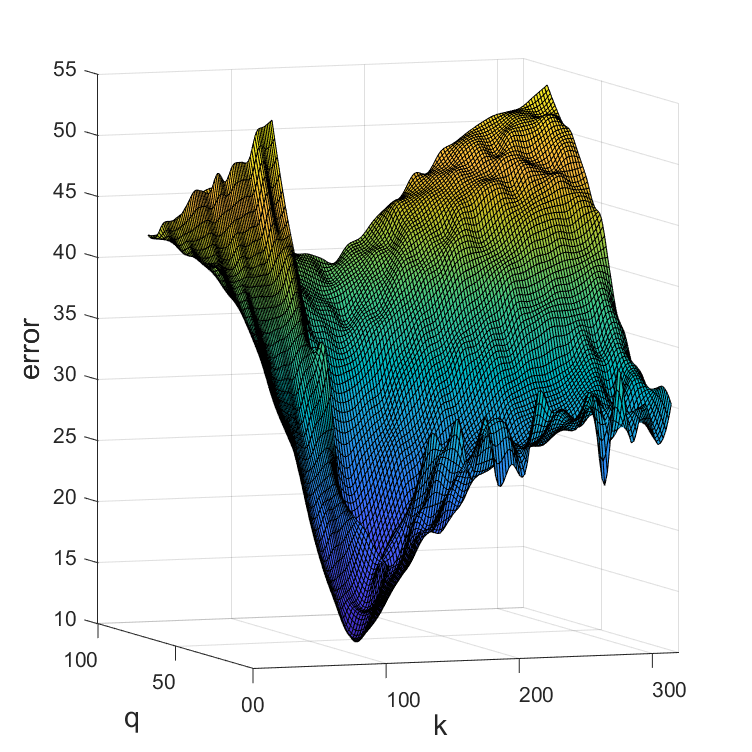}\hspace{0.2in} \includegraphics[width=0.5\textwidth]{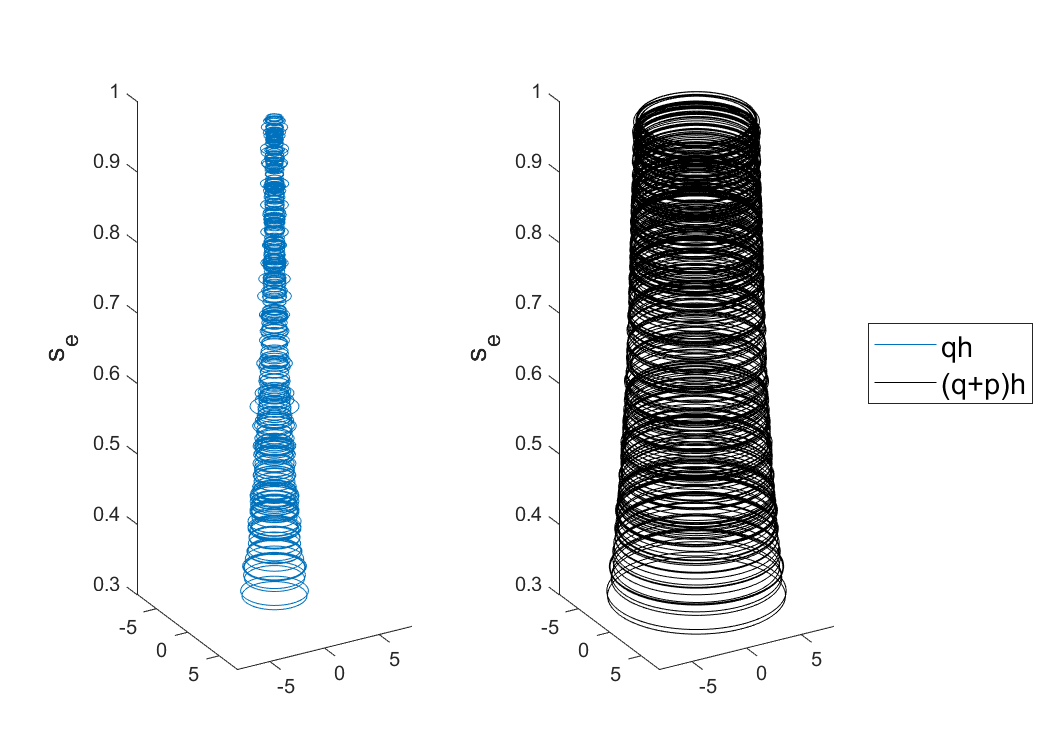}
    \caption{Left panel: Error landscape calculated over a uniform mesh in the $[20,320]\times[5, 85]$ region of the ({\tt k}, {\tt q}) parameter space. The surface is plotted using cubic interpolation over the actual data. Note the larger variability of $E$ for very small values of {\tt q}. Right panel: PSO results from 500 runs (as usual, outliers associated with a large error are removed; see main text for details). The optimal trap efficiency {\tt s}$_{\tt e}$ is plotted as a function of the optimal inner radius {\tt q} $\cdot$ {\tt h} (blue circles, left panel) and of the optimal outer radius ({\tt q}+{\tt p}) $\cdot$ {\tt h} (black circles, right panel) of each capture site}
    \label{fig:surf-Cairns-PSO-pqCircvseff}
\end{figure}

For the grid search, we run forward simulations over a uniform sampling of {\tt k} and {\tt q} while setting {\tt p} = 0 and ${\tt s}_{\tt e} = 100\%$. We take ({\tt k}, {\tt q}) in the range $[20,320]\times[5, 85]$, with a total of 527 parameter pairs for each (31 values for {\tt k} and 17 values for {\tt q}). The resulting error landscape, with the error defined in Equation (\ref{eq:Error2}), is displayed on the left panel of Figure \ref{fig:surf-Cairns-PSO-pqCircvseff}. It shows a clear minimum, around which the initial conditions for the PSO step of the method are distributed. 

As shown in the main text, the results of the PSO step confirm that the optimal values of {\tt k} follow a normal distribution. The pattern followed by the optimal values of {\tt q}, {\tt p} and ${\tt s}_{\tt e}$ is visualized on the right of Figure \ref{fig:surf-Cairns-PSO-pqCircvseff}. As in the Hainan study, when the value of ${\tt s}_{\tt e}$ is high, the values of {\tt q} and {\tt p} are low, and vice versa.

\section{Properties of MDT, TC, and ATC estimates}
\label{SP:ATCM_prop}
This section uses synthetic data to explore how the values of $D_{MDT}$, $D_{TC}$, and $D_{ATC}$ are affected by the MRR experimental setup. A comparison with the results of the RDA-PSO method (based on 20 independent runs) is also provided.

\subsection{Uniform trap density: Sensitivity of TC estimates to experimental conditions} \label{sec:DTCM_sen}

\begin{figure}[hbtp]
\centering
    \includegraphics[width=0.42\textwidth]{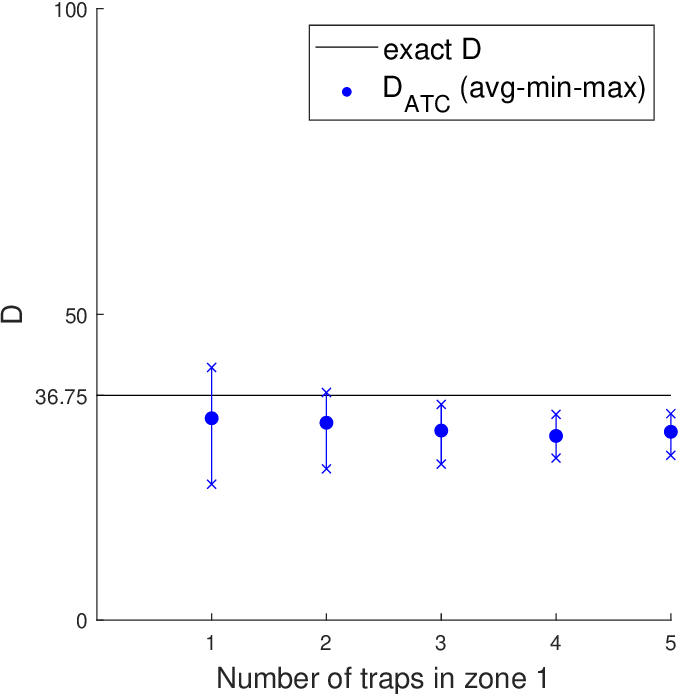}\hspace{0.4in} \includegraphics[width=0.42\textwidth]{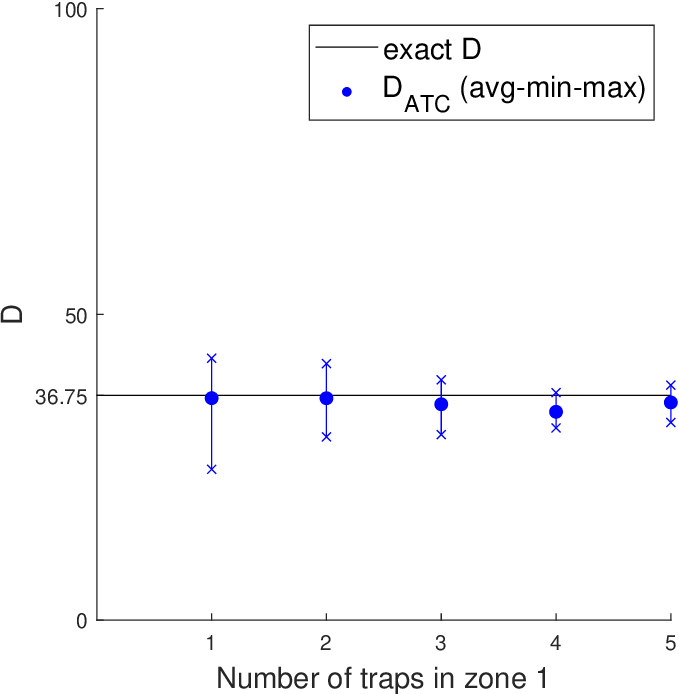}
    \caption{Estimates of $D_{TC}$ for different trap densities. The $x$-axis shows the number of traps in the first zone, $nT_1$; the number of traps in the other zones is an odd multiple of $nT_1$, per the definition of $nT_j$. For each $nT_1$ value, there are 100 sets of randomized trap locations. The large blue dot is the average value and the crosses mark the maximum and minimum estimated values of $D_{TC}$. Parameters are {\tt k} = 35, {\tt q} = 20, {\tt p} = 0, and $N=10,000$. Trap efficiency is 3\%. The horizontal line represents the exact value of the diffusion coefficient $D$. Left panel: Temporal recapture ratios (see Equation \eqref{eq:time_pct} of the main text) are calculated daily. Right panel: Temporal recapture ratios are calculated hourly} \label{fig:DTC_randomization}
\end{figure}

\begin{figure}[hbtp]
\centering
    \includegraphics[width=0.42\textwidth]{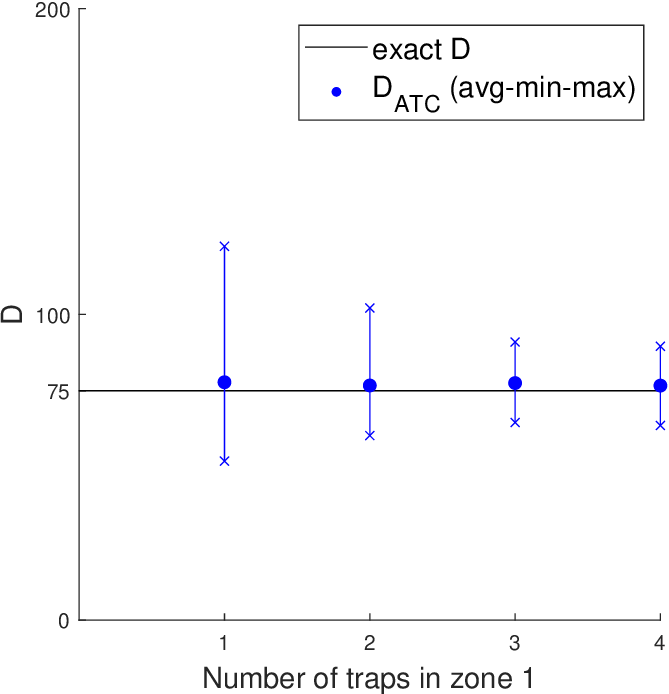}\hspace{0.4in} \includegraphics[width=0.42\textwidth]{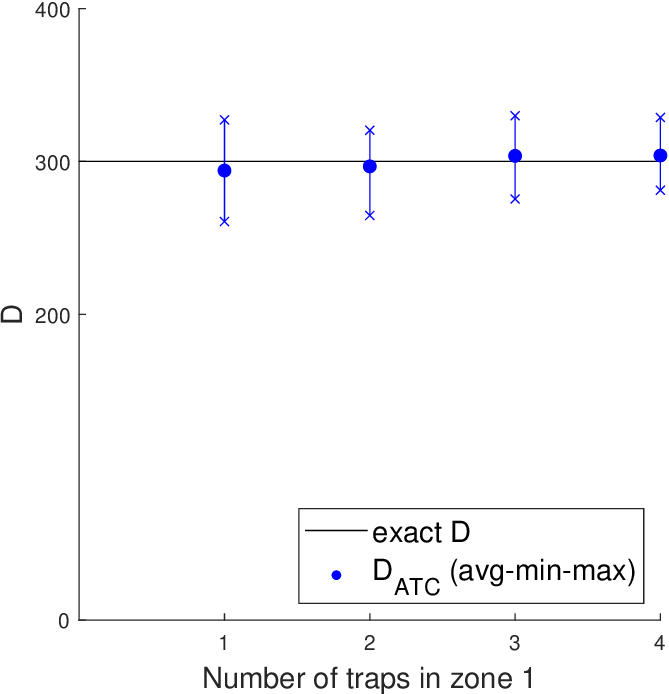}
    \caption{Similar to the right panel of Figure \ref{fig:DTC_randomization}, but for 50 sets of randomized trap locations, larger recapture zones ($r_{zone} = 50$ m), and different values of {\tt k}. Hourly time recapture ratios are considered here and the traps have 3\% efficiency. Left panel: {\tt k} = 50. Right panel: {\tt k} = 100}
    \label{fig:DTC_randomization_large_k_hourly}
\end{figure}

Unless otherwise mentioned, all of the virtual MRR experiments described below are conducted with {\tt k} = 35 and {\tt q} = 20. The setup is similar to that of the Hainan center study, with 6 concentric zones separated by $r_{zone} = 15$ meters. Traps are uniformly distributed in each zone, in numbers proportional to their area. Specifically, the area of the $j$-th zone, $j \ge 1$, is 
\[
A_j = \pi [(j r_{zone})^2 - ((j-1) r_{zone})^2] = (2j-1) \pi r_{zone}^2 = (2j-1) A_1,
\]
and the number of traps in zone $j$ is given by $nT_j =  (A_j/A_1)\, nT_1 = (2j-1)\, nT_1$, where $nT_1$ is an integer at least equal to one. Each virtual experiment runs the forward model over a period of 10 days, with $N=10,000$ agents, {\tt p} = 0, and a trap efficiency {\tt s}$_{\tt e} = $ 3\%. In the simulations presented here, this means that once an agent reaches the collection region of a trap, there is a 3\% chance of it being caught. If not captured but still within the collection region after one random walk step, this chance is given again, until the agent is either captured or moves away from the trap. Associated recapture ratios are recorded and used to estimate $D_{TC}$ by finding the value of $D$ that simultaneously optimizes $Q_{tot}$ for all of the zones (see Equation \eqref{eq:Q_tot_model} of the main text). Since the number of traps is proportional to the area of each zone, $D_{TC} = D_{ATC}$ in the present situation. We create 100 different sets of random trap locations over which forward simulations are run and we report the average value of $D_{TC}$ as well as its maximum and minimum. This experiment is repeated for different trap densities, or equivalently values of $nT_1$, the number of traps in zone 1 (horizontal axis of Figures \ref{fig:DTC_randomization} and \ref{fig:DTC_randomization_large_k_hourly}). 

The left panel of Figure \ref{fig:DTC_randomization} shows that the average value of $D_{TC}$ (large dots) is robust to changes in trap density, as long as the number of traps in each zone is proportional to its surface area. However, there is wider variation around the average of $D_{TC}$ (the crosses represent the minimum and maximum estimates over 100 sets of trap locations) when fewer traps in total are deployed. Although time-corrected estimates of $D$ are, on average, consistent across trap density and location, they still under-estimate the true value of $D$, which corresponds to {\tt k} = 35 and is represented by the horizontal line in Figure \ref{fig:DTC_randomization}.

Estimates of $D_{TC}$ may be improved by increasing the frequency at which temporal recapture ratios (see Equation \eqref{eq:time_pct} of the main text) are evaluated. Since in the simulations walkers are captured every hour, the most accurate temporal recapture ratios available are hourly. Using such ratios leads to average values of $D_{TC}$ that are closer to the actual value of $D$, as shown in the right panel of Figure \ref{fig:DTC_randomization}. As before, increasing the trap density reduces uncertainty around the mean.

We checked that similar results are obtained for smaller ({\tt k} = 15) and larger ({\tt k} = 50) values of {\tt k}, indicating that {\em for equal trap densities across zones}, the time-corrected model produces correct estimates of $D$ using hourly recapture ratios, regardless of the average dispersal rate. (Recall that throughout this section the time-corrected and time-and-area-corrected models are the same, that is, $D_{ATC}= D_{TC}$). Finally, Figure \ref{fig:DTC_randomization_large_k_hourly}, which summarizes the results of 50 sets of randomized trap locations for different values of $nT_1$, shows that estimates of $D_{TC}$ from hourly time ratios remain robust even if the width $r_{zone}$ of each zone is increased, in this case to $r_{zone}=50$ meters, as in the Cairns study. Trap efficiency is kept at 3\% but $N$ is adjusted to $N=20,000$ to accommodate the larger study area (corresponding to 4 concentric zones). The parameter {\tt q} is fixed at {\tt q} = 30 and simulations are run for {\tt k} = 50 (left panel of Figure \ref{fig:DTC_randomization_large_k_hourly}) and {\tt k} = 100 (right panel of Figure \ref{fig:DTC_randomization_large_k_hourly}). Note that, at this scale, a slight over-estimate of $D$ is as likely as a slight under-estimate. 

\subsection{Heterogeneous trap density: Limitations of ATC and MDT estimates} \label{sec:DATCM_sen}

\begin{figure}[hbtp]
\centering
    \includegraphics[width=\textwidth]{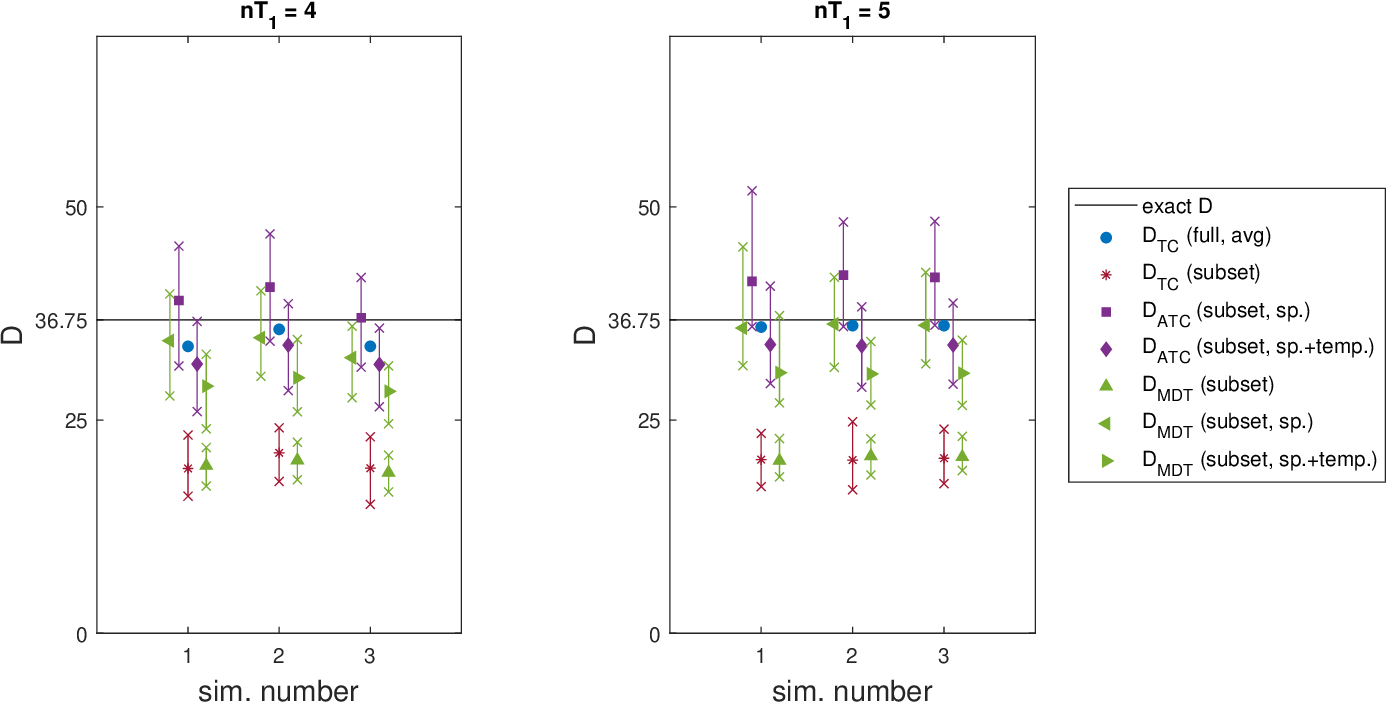}
    \caption{Diffusion coefficient $D_{TC}$ estimates for 6 randomized full set of traps (blue dots) and 50 randomized subsets of these full sets (dark red stars). Estimates of $D_{ATC}$ with corrected spatial ratios (purple squares) and both spatial and temporal ratios (purple diamonds) are also shown. The green triangles correspond to $D_{MDT}$ values, with upper, left, and right triangles using uncorrected, corrected spatial, and corrected spatial and temporal ratios, respectively.  In each case, the crosses indicate the minimum and the maximum estimates. The parameters are set at {\tt k} = 35, {\tt q} = 20, {\tt p} = 0, and ${\tt s}_{\tt e} = 3$~\%, with $N = 10,000$ walkers over 6 zones with $r_{zone} = 15$ meters. The horizontal black line indicates the exact value of $D$ obtained from the fixed {\tt k} parameter}
    \label{fig:HPC-D-Virtual-Trapdensity-4-5-k-35-q-20-subset-ByHr-tcorr}
\end{figure}

When trap density is not uniform across capture zones, the time-corrected model needs to be adjusted with the correction factor of Equation \eqref{eq:CF} \citep{LMJ81} or alternatively \eqref{eq:CorrectionMatrixCount}, thereby leading to the area-and-time-corrected model. To study the relation between the two models, we create 2 virtual MRR experiments with uniform trap distribution (each over 6 zones with $r_{zone} = 15$ meters), the first one with $nT_1 = 4$ and the second with $nT_1 = 5$. We refer to each set of traps as a ``full set'' and randomly create 3 such full sets, for both $nT_1 = 4$ and $nT_1 = 5.$ Then we randomly choose 50 subsets of each full set to match the trap count of the Hainan center experiment, which has 4, 7, 6, 2, 0, and 1 traps in zones 1 through 6 respectively. We run simulations over each subset with $N = 10,000$ walkers released from the center and parameters set at {\tt k} = 35, {\tt q} = 20, {\tt p} = 0, and ${\tt s}_{\tt e} = 3$~\%. Each simulation is run over a period of 10 days and hourly time recapture ratios are collected. Figure \ref{fig:HPC-D-Virtual-Trapdensity-4-5-k-35-q-20-subset-ByHr-tcorr} shows 3 sets of data for each synthetic study, which include: the average of 10 estimates of $D_{TC}$ over the full set of traps (blue dots), the average of $D_{TC}$ estimates over the 50 subsets (dark red stars), the average of $D_{ATC}$ estimates over the 50 subsets when temporal ratios are not corrected (purple squares), and the average of $D_{ATC}$ estimates over the 50 subsets with corrected temporal ratios (purple diamonds), as well as estimates of $D_{MDT}$ using uncorrected ratios (green up triangles), corrected spatial ratios (green left triangles), and corrected spatial and temporal ratios (green right triangles). In each case, we also indicate the minimum and the maximum of the corresponding estimates of $D$.

As expected, $D_{TC}$ estimates obtained from the 50 subsets (dark red stars) are low, since they use uncorrected ratios. Correcting both temporal and spatial ratios (purple diamonds) leads to $D_{ATC}$ estimates that are closer to the exact value of $D$, obtained from \eqref{eq:DTrue_k} with {\tt k} = 35. However, in all cases, the average value of $D_{ATC}$ is an under-estimate of $D$, whereas individual values are either over- or under-estimates, even though recapture ratios were calculated hourly.

Estimates of $D_{MDT}$ based on uncorrected ratios are low and comparable to those of $D_{TC}$ over each subset. Correcting spatial ratios leads to averaged values of $D_{MDT}$ that are comparable to the actual value of $D$. Surprisingly, correcting both temporal and spatial ratios typically makes $D_{MDT}$ an underestimate of $D$. In fact, $D_{MDT}$ seems to give lower estimates than $D_{ATC}$ with the same input data.

\subsection{Comparison with RDA-PSO estimates} \label{sec:DATCM_wPSO}
\begin{figure}[hbtp]
\centering
    \includegraphics[width=0.8\textwidth]{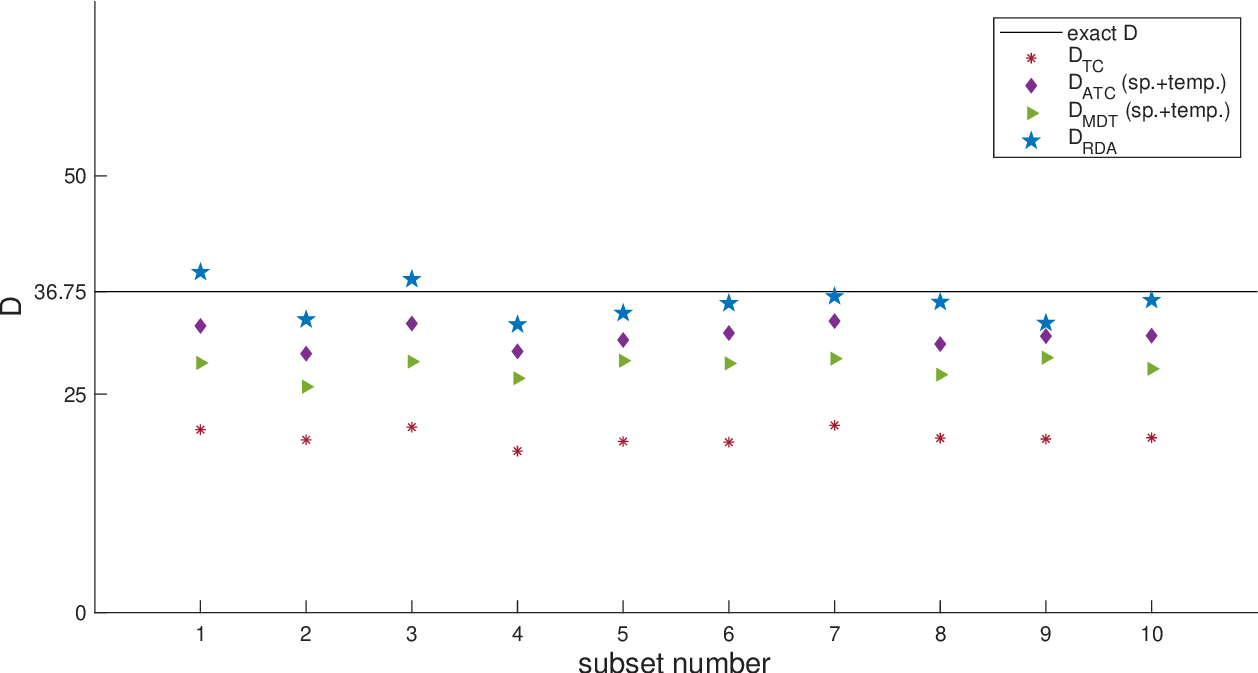}
    \caption{Comparison of the diffusion coefficient estimates for the four approaches introduced in this article, for 10 different subsets of randomly generated trap locations. The estimates for $D_{MDT}$ and $D_{ATC}$ are computed with both spatial and temporal corrections. The $D_{RDA}$ estimate is calculated from 20 PSO runs after removing outliers. The parameters are set at {\tt k} = 35, {\tt q} = 20, {\tt p} = 0, and ${\tt s}_{\tt e} = 3$~\%, with $N = 3,000$ walkers over 6 zones with $r_{zone} = 15$ meters. The horizontal black line indicates the exact value of $D$ obtained from the fixed {\tt k} parameter}
    \label{fig:HPC-D-Virtual-Trapdensity-4-k-35-q-20-subsets-10-ByHr-wPSO}
\end{figure}

This section compares estimates returned by the MDT, TC, ATC, and RDA-PSO models for synthetic experiments with trap configurations similar to those of Section \ref{sec:DATCM_sen}. We randomly generate 10 subsets of trap locations from the full set used for Simulation 1 of the previous section, for which $nT_1=4$. Input data consist of the recapture ratios of $N = 3,000$ walkers, with parameters set at {\tt k} = 35, {\tt q} = 20, {\tt p} = 0, and ${\tt s}_{\tt e} = 3$~\%, over 6 zones with $r_{zone} = 15$ meters. For each subset, we calculate $D_{MDT}$, $D_{TC}$, $D_{ATC}$, using both the spatially and temporally corrected ratios (for the MDT and ATC estimates). To calculate $D_{RDA}$, we obtain optimal values of {\tt k} from 20 PSO runs with $N_p = 15$ particles and $N_g = 5$ generations since this is a 2-parameter optimization, remove outliers, find the average value of the remaining {\tt k}'s, and substitute the result in equation (\ref{eq:DTrue_k}). Figure \ref{fig:HPC-D-Virtual-Trapdensity-4-k-35-q-20-subsets-10-ByHr-wPSO} compares the resulting estimates with the known value of $D$, for each of the 10 subsets of trap configurations. Minimum and maximum values of the signed relative errors for each of the 4 methods are listed in Table \ref{tab:DATCM-wPSO}. These results show that, in situations where the true value of the diffusion coefficient is known, the RDA-PSO outperforms the other three methods. In addition, the RDA-PSO estimate has a relative error of less than 11\% when 20 PSO simulations are used.

\begin{table}[hbtp]
\centering
    \caption{Minimum and maximum relative errors for the estimates shown in Figure \ref{fig:HPC-D-Virtual-Trapdensity-4-k-35-q-20-subsets-10-ByHr-wPSO}. Errors are calculated with respect to the known value of $D$, equal to 36.75 m$^2/$day.}
    \label{tab:DATCM-wPSO}
    \begin{tabular}{lcc}
        \toprule%
        & minimum relative error  & maximum relative error \\
        &  (from the exact $D$) & (from the exact $D$) \\
        \midrule
          $D_{MDT}$ (sp.+temp.) &  $-29.64\%$ & $-20.6\%$\\
          $D_{TC}$ & $-49.7\%$ & $-41.69\%$\\
          $D_{ATC}$ (sp.+temp.) & $-19.35\%$ & $-9.21\%$\\
          $D_{RDA}$ &  $-10.26\%$ & $6.05\%$\\
        \botrule%
    \end{tabular}
\end{table}

\section{Effect of trap density on recapture ratios}
\label{SP:TrapDensity}
\begin{figure}[hbtp]
\centering  
    \includegraphics[width=0.45\textwidth]{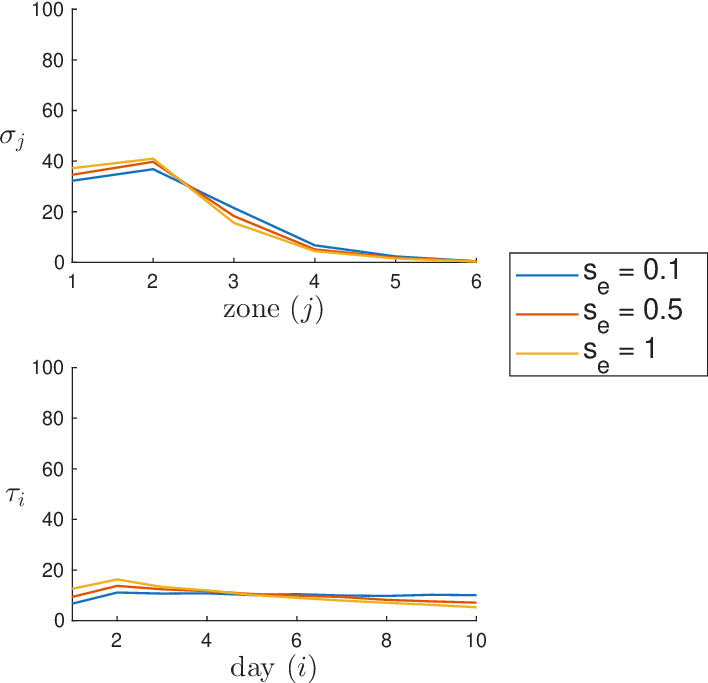}\hspace{0.2in} \includegraphics[width=0.45\textwidth]{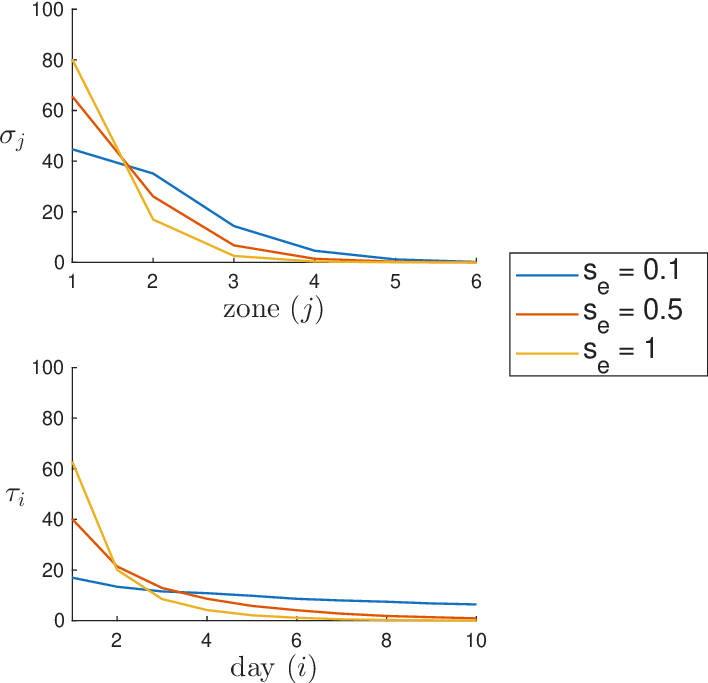}
    \caption{Recapture ratios, $\sigma_j$ (top plot) and $\tau_i$ (bottom plot), corresponding to virtual experiments with different trap efficiency, {\tt s}$_{\tt e}$. Traps are distributed uniformly according to zone areas, with 6 concentric zones $15$ apart. Left panel: The first zone has 1 trap (low trap density). Right panel: The first zone has 4 traps (high trap density)}
    \label{fig:Pct-Virtual-Trapdensity-1-4-k-35-q-20-captEff-0p1-1}
\end{figure}

Recapture ratios, $\tau_i$ and $\sigma_j$, are sometimes affected by the value of trap efficiency,  {\tt s}$_{\tt e}$, depending on how many traps are used in an experiment. If the trap density is low, recapture ratios remain robust to changes in {\tt s}$_{\tt e}$. However, if the trap density is high, recapture ratios change depending on the value of {\tt s}$_{\tt e}$. This is illustrated in the following virtual experiment. We consider 6 concentric zones (separated by $r_{zone} = 15$ meters, as in Appendix \ref{SP:ATCM_prop}) and a uniform trap density. In the first scenario, we place 1 trap in the first zone, that is, $nT_1 = 1$; whereas in the second scenario, $nT_1 = 4$. We release $N=10,000$ mosquitoes and for every forward run, we vary the trap efficiency to see how the recapture percentages change. 
The left panel of Figure \ref{fig:Pct-Virtual-Trapdensity-1-4-k-35-q-20-captEff-0p1-1} shows that when less traps are used ($nT_1 = 1$), the percentages are close to each other, no matter the value of {\tt s}$_{\tt e}$. On the other hand, the right panel of Figure \ref{fig:Pct-Virtual-Trapdensity-1-4-k-35-q-20-captEff-0p1-1} shows that the recapture percentages vary considerably with {\tt s}$_{\tt e}$ when the number of traps is high.  Since the traps are uniformly distributed, having 4 traps in zone 1 means there are 144 traps in the entire field of study. In that case, the right panels of Figure \ref{fig:Pct-Virtual-Trapdensity-1-4-k-35-q-20-captEff-0p1-1} indicate that most of the walkers are captured on the first two days and in the first two zones when {\tt s}$_{\tt e}$ is high, suggesting the associated MRR experiment is badly designed. In other words, more is not better in this case, and the number of traps used for a MRR experiment should be chosen so that the spatial and temporal ratios remain non-negligible in most of the zones, for the entire duration of the experiment.

\end{appendices}

\newpage

\end{document}